# Health Impacts of Public Pawnshops in Industrializing Tokyo

Tatsuki Inoue*

Abstract

This study is the first to investigate whether pawnshops, financial institutions for low-income populations, have contributed to the decline in mortality in the early twentieth century. Using ward-level panel data from Tokyo City, this study revealed that the popularity of public pawnshops was associated with a 4% and 5% decrease in infant mortality and fetal death rates, respectively, during 1927–1935. The historical context implies that the potential channels of the relationships were improving nutrition and hygiene and covering childbirth costs. Moreover, a cost-effectiveness calculation highlighted that the establishment of public pawnshops was a cost-effective public investment for better public health. Contrarily, for-profit private pawnshops showed no significant association with health improvements.

---

* School of Commerce, Meiji University, Faculty Office Building, 1-1, Kanda-Surugadai, Chiyoda-ku, Tokyo, 101-8301, Japan. E-mail: tatsukiinoue@meiji.ac.jp. I am grateful to Kota Ogasawara and Shonosuke Sugasawa for their invaluable comments. This work was supported by Grant-in-Aid for Early-Career Scientists [Grant Number: 22K13440]. There are no conflicts of interest to declare. All errors are my own.

## 1. Introduction

The economic and demographic history literature has widely examined the dramatic decline in infant mortality that Japan and Western countries experienced in the early twentieth century. As Goldin (2024) notes, public health interventions played a crucial role in this transformation.[1] Previous literature has also highlighted the contributions of personalized health interventions, national health insurance programs, and cash transfers in curbing mortality rates (Bauernschuster et al., 2020; Bhalotra et al., 2017; Bowblis, 2010; Galofré, 2020; Moehling and Thomasson 2014; Ogasawara and Kobayashi, 2015; Winegarden and Murray, 1998; 2004; Wüst, 2012). However, despite the prevalence of liquidity constraints in both historical and contemporary contexts, the impact of household access to credit, especially for the poor, on survival outcomes has received far less attention. Easing such constraints could help vulnerable families smooth consumption during income shocks, maintain adequate nutrition, and pay for essential health needs. However, whether improved credit access, independently of increases in household income, contributed to the decline in early-life mortality remains unclear.

[1] More broadly, Ogasawara (2024a) and Schneider (2025) provide comprehensive reviews on child health and stunting.

To fill this knowledge gap, the current study examines the unique case of public pawnshops in Tokyo City during the 1920s and 1930s. During this period, many low-income households in Tokyo faced acute liquidity constraints due to their insufficient access to credit. In response, the Tokyo City government established public pawnshops designed to help the urban poor, unlike traditional private counterparts, which operated on a for-profit basis. By providing low-interest, small loans with priority given to accessibility for the poor, these institutions may have eased the liquidity constraints faced by low-income households, thereby allowing them to maintain essential expenditures on food, medical care, and hygiene. With a particular focus on infant and fetal mortality rates, this study investigates whether such interventions contributed to improvements in early-life health.

Using newly digitized ward-level data in Tokyo City from 1927 to 1935, this study examines the impact of increases in public pawnshop loans on early-life mortality. Employing fixed-effects models to control for unobserved ward- and year-specific characteristics, this study finds that greater access to public pawnshops is associated with a 3.69% and 4.89% reduction in infant mortality and fetal death rates, respectively, over the study period. Cost-effectiveness analysis further demonstrates that public investment in these institutions was highly efficient: for every 1,023 yen (approx. $505 in 1930 dollars or $6,593 in 2010 dollars), one additional infant death could have been prevented. These effects likely reflect improved nutrition and hygiene, as well as the ability to cover pregnancy and childbirth expenses, which were made possible by easier access to small loans. By contrast, private pawnshops, despite their large-scale operations, did not exhibit any significant relationship with health outcomes.

This study makes three key contributions. First, to the best of the author's knowledge, this study is the first to empirically demonstrate that providing low-interest loans to low-income households contributed to a reduction in early-life mortality in the early twentieth century. While numerous studies, including those in Japan, have established the health benefits of public health interventions such as improvements in water supply and sewage systems (Alsan and Goldin, 2019; Anderson et al., 2022; Cutler and Miller, 2005; Ferrie and Troesken, 2008; Inoue and Ogasawara, 2020; Ogasawara and Matsushita, 2018, 2019; Ogasawara et al., 2018), no previous research has investigated the effect of relaxing liquidity constraints for economically vulnerable populations, specifically through the provision of low-interest loans. By clarifying this relationship, the present study provides evidence for a previously underexplored channel in the historical debate on early-life health.

Second, the current study offers insight into the broader significance of risk-coping strategies for vulnerable households in historical contexts. Much research has demonstrated that formal and informal financial resources, such as precautionary savings, lending institutions, transfers, and charity, help low-income individuals smooth their consumption during the prewar period (Horrell and Oxley, 2000; James and Suto, 2011; Kiesling, 1996; Saaritsa, 2008, 2011; Scott and Walker 2012). Furthermore, in a closely related study, Ogasawara (2024b) determined that working-class households in early 1920s Osaka used pawnshops to mitigate idiosyncratic shocks. However, these studies did not address whether such risk-coping behavior led to tangible improvements in living standards. The findings in the present study advance our understanding of the health consequences of risk-coping strategies in times of economic hardship.

Finally, this research encourages a reconsideration of the role of pawnshops as an alternative financial service from a historical perspective. Although the evidence on the advantages and drawbacks of modern fringe banking remains inconclusive (Bhutta, 2014), historical studies have documented that private pawnshops served as key financial resources in Japan and Ireland during unforeseen disasters such as pandemics and famines (Inoue, 2021; McLaughlin, 2022). This study adds positive evidence to the debate by showing that public pawnshops, which were characterized as charitable organizations, alleviated financial distress and improved public health even outside of disasters. Moreover, these institutions achieved both cost-effectiveness and financial sustainability through modest interest payments. These findings suggest that, with appropriate institutional design, alternative financial services like pawnshops can effectively promote financial inclusion and serve as social safety nets.

## 2. Historical Background

### 2.1. Development of Pawnshops

Pawnshops in Japan have a rich history as traditional financial institutions providing secured loans using relatively inexpensive items as collateral. Focusing on collateral value allows pawnshops to reduce the transaction costs derived from information asymmetry and offer loans without credit checks. These small pawn loans were essential financial resources for low-income households, especially for those without access to banking services.

Private for-profit pawnshops have existed in Japan since the thirteenth century and spread across the country, reaching 17,000 in 1924 (Social Welfare Bureau, 1926a).[2] These pawnshops are primary lending institutions for low-income households because of their lower interest rates compared with those charged by loan sharks.[3] However, these rates are not sufficiently low and are deemed to hurt low-income individuals (Tokyo Institute for Municipal Research, 1926; Shibuya et al., 1982).

With the growing social problem of poverty around 1910, the Home Ministry, Ministry of Finance, and Bank of Japan investigated public pawnbroking systems in several European countries, including Italy, France, Belgium, the Netherlands, Germany,

---

[2] The first writing for pawnbroking can be traced back to the Taiho Code, established in 701 (Bank of Japan, 1913). This pawnbroking system is believed to have partially originated in China (Tokyo Pawnshop Association, 1934).

[3] In poor areas of prewar Japan, interest is charged even on loans from personal networks, including relatives, acquaintances, and friends. These interest rates are often comparable to those of loan sharks (Kojima, 2021). According to the Tokyo Prefecture Department of Academic Affairs (1935), approximately 30% of such loans have a monthly interest rate of 5%, while 25% have rates of 10% or higher.

Austria, and Spain (Tokyo Institute for Municipal Research, 1926). In 1912, the first public pawnshop opened in Hosoda Village, Miyazaki Prefecture, to help low-income villagers. These movements sparked debates over private and public pawnshops. For example, Toyohiko Kagawa, a social activist, criticized private pawnshops' interest rates as disadvantageous to the poor compared with that of European pawnshops, as they charged higher interest rates for smaller loans (Kagawa, 1915).[4] Conversely, the owners of private pawnshops claim that their interest rates are not unreasonably high (Ogasawara, 1913).

In response to the public's demand for pawnshops to serve as social enterprises and fierce opposition from private pawnbrokers, the government conducted many surveys and decided to allow both private and public pawnshops (Shibuya et al., 1982). Consequently, the Public Pawnshop Law (*Koeki shichiya ho*) was established in 1927 and Pawnshop Control Law (*Shichiya torisihmari ho*) for private pawnshops was retained. This marked a turning point in the development of public pawnshops; with only 31 public pawnshops in 1925 and 81 in 1927, their number grew rapidly to 1,079 by 1935.

[4] He was one of the most famous Japanese in the world at that time. See Shaffer (2013) for more detail.

## 2.2. Pawnshops in Tokyo

Pawnshops are primarily located in urban areas such as Tokyo. Figure 1 illustrates the number of private and public pawnshops in Tokyo between 1920 and 1935. In 1920, over 1,200 private pawnshops operated; however, public pawnshops had not been established. However, the Great Kanto Earthquake of 1923 caused a significant decrease in the number of private pawnshops, with only 500 operations remaining shortly after the disaster.[5] The Social Welfare Bureau (1926b) notes that a complete recovery from earthquakes is difficult. Although smaller pawnshops borrowed funds from larger ones when they are short of cash, the financial instability caused by the earthquake made this impossible. It is challenging for small pawnbrokers with little credit to run their businesses using funds alone. Additionally, warehouses for storing pawn items are essential for pawnshops. However, rebuilding warehouses destroyed by the earthquake

[5] The Great Kanto Earthquake was a major earthquake with a moment magnitude of 7.9 that struck the Kanto region of Japan, particularly Tokyo and Yokohama, on September 1, 1923. The total number of casualties and missing people was estimated to be over 100 thousand. This earthquake had not only short-term but also long-term impacts on economies and human capital (Ogasawara, 2022; Okazaki et al., 2019).

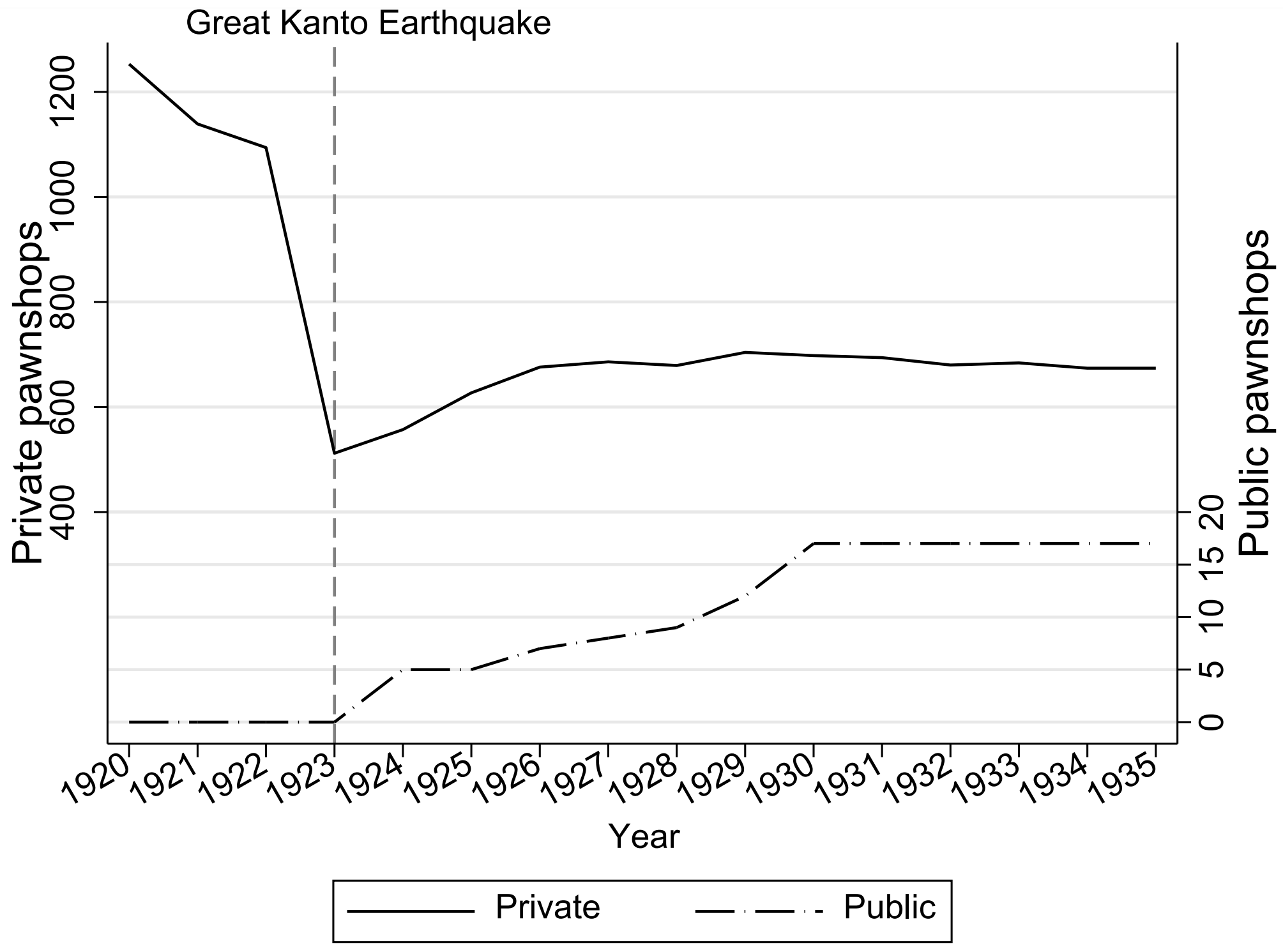

Figure 1: Number of Pawnshops in Tokyo City

Note: The number of public pawnshops indicate the number established by Tokyo City in the old area. The decline in the number of private pawnshops in 1923 was due to the Great Kanto Earthquake of 1923. Sources: Social Welfare Bureau (1926b, pp. 6–7); Tokyo City Social Welfare Bureau (1928–1935); Tokyo Prefecture (1926–1937); Tokyo Prefecture Department of Academic Affairs (1935).

required a significant amount of money. Eventually, the number of private pawnshops returned to approximately 700 by 1926 and remained stable until 1935, indicating a stable management and availability of sufficient loans to borrowers.

In contrast to the decrease in private pawnshops, Tokyo established five public pawnshops in Asakusa, Fukagawa, Honjo, Kyobashi, and Shimoya wards in 1924 in response to the increased requirement for lending institutions among economically disadvantaged people following the earthquake.[6] The number of public pawnshops rose to 8 in 1927 and 17 in 1930 supported by government subsidies and donations for earthquake relief (Tokyo Prefecture Department of Academic Affairs, 1935). This increase reflects the growing demand for such institutions.[7]

Geographical concentration of pawnshops is important when considering their role in serving the financial needs of low-income communities. Pawnshops are primarily situated in poor areas and provide households with limited financial resources access to small short-term loans. Figure 2 presents the spatial distribution of annual income per capita and pawnshops across Tokyo (Old City area).[8] Figure 2a shows that Shimoya,

[6] Although Tokyo City expanded its area in 1932, the newly incorporated areas are considered rural compared with the old areas, as they have been towns and villages before integrating in Tokyo. To maintain consistency, this study focuses on 15 wards in the old area. In addition, it only treats public pawnshops established by Tokyo City owing to data limitations.

[7] In addition to public pawnshops, Tokyo provided various social welfare services, such as social workers, municipal houses, public bathhouses, cheap lodging houses, eating places, employment security offices, vocational aid centers, maternity hospitals, dispatched midwives for home birth, day nurseries, shelters for children, and settlement houses (some of which started from 1929). Among these, pawnshops are expected to be the most helpful by the authorities (Tokyo City Social Welfare Bureau, 1930).

[8] The income per capita is derived from the income survey in Tokyo City Office (1933a). In 1930, the Statistical Division of Tokyo City surveyed 4,227 business corporations, 61,112 taxpayers and their family

Asakusa, Honjo, and Fukagawa wards are the poorest areas, with annual incomes per capita ranging from 296 to 320 yen. This income level is approximately 50% of that of the wealthiest areas, such as Kojimachi, Azabu, Akasaka, and Nihombashi wards. Figure 2b shows that public pawnshops are established in low-income wards. While there are more than three public pawnshops in each eastern ward, the more affluent wards have only one or no public pawnshops.

Figures 2c and 2d, which present the distribution of private pawnshops in 1921 and 1933, respectively, illustrate a similar pattern to that of public pawnshops: a higher concentration in the economically challenged eastern wards. The distribution of private pawnshops remained relatively constant between 1921 and 1933 even in the face of the significant destruction caused by the Great Kanto Earthquake. This persistence suggests members, and 1,802 individuals whose income is less than the criterion for paying income tax. This survey asked respondents about their income and its sources in detail. Therefore, the income per capita includes the income from employment and other sources, such as interest and stock.

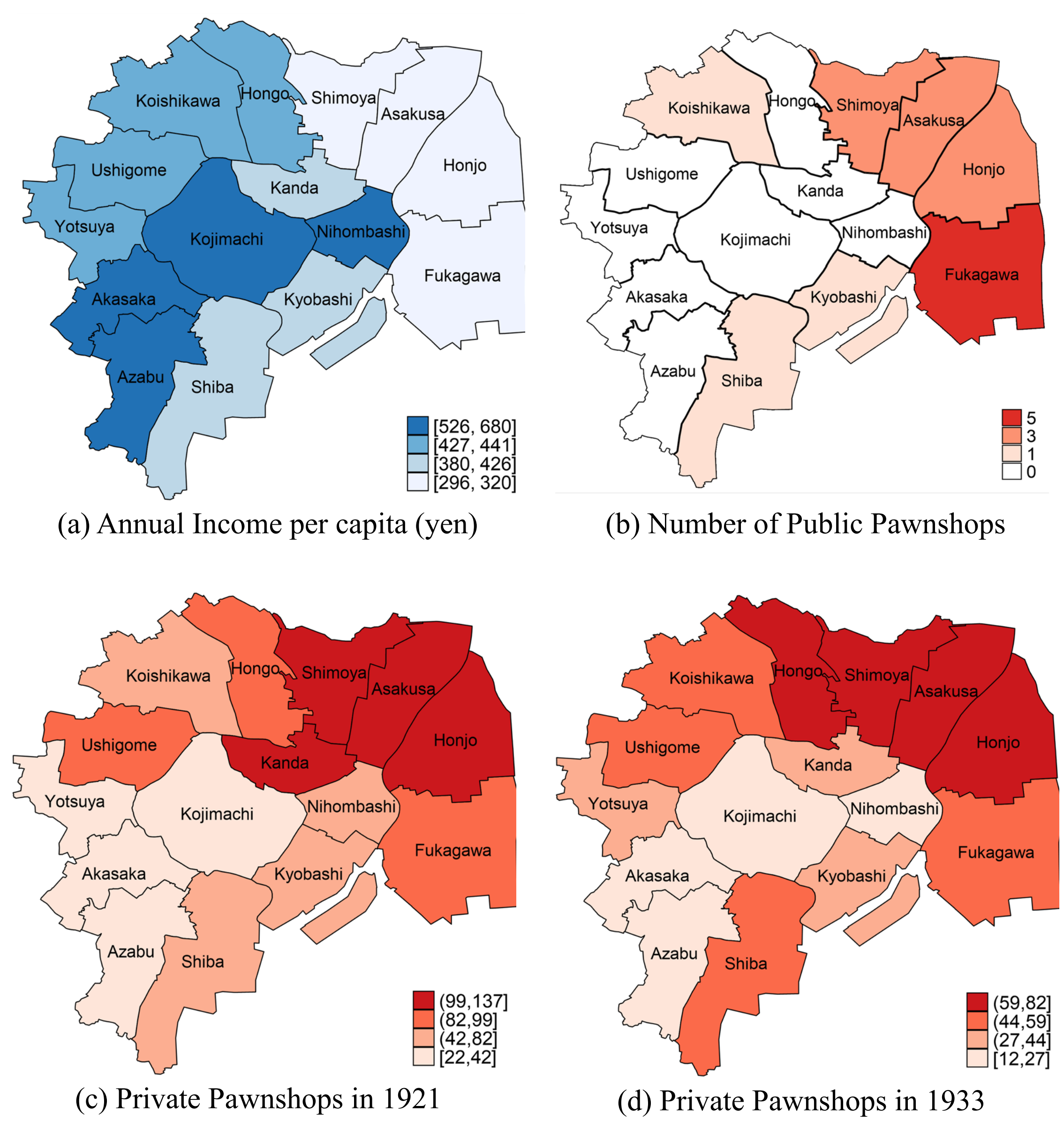

(a) Annual Income per capita (yen)

(b) Number of Public Pawnshops

(c) Private Pawnshops in 1921

(d) Private Pawnshops in 1933

Figure 2: Spatial Distribution of Income and Pawnshops across Tokyo

Note: Figure 2a shows the average annual income per capita in 1933. It also includes the income from interest and stock. Figure 2b shows the spatial distribution of public pawnshops in 1933. Figures 2c and 2d show the spatial distribution of private pawnshops in 1921 and 1933, respectively. This data covers only the Old City area. Sources: Social Welfare Bureau (1926b); Tokyo City Office (1933); Tokyo Prefecture (1935); Tokyo Prefecture Department of Academic Affairs (1935).

high demand for pawnshops among low-income households and their role in providing financial services.

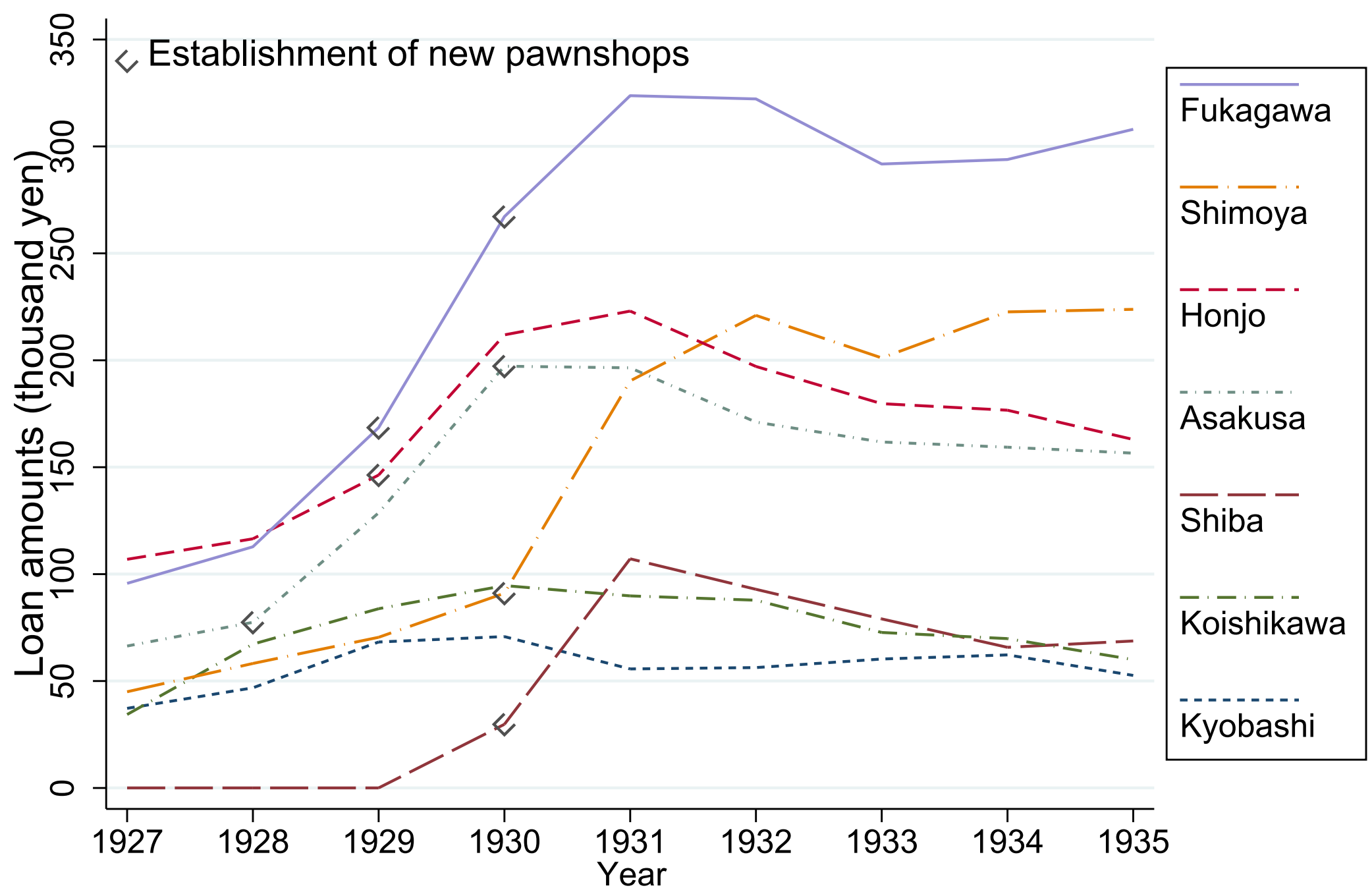

Figure 3: Loan Amounts of Public Pawnshops in Different Wards

Note: Loan amounts are the real values adjusted by the wholesale price index based on December 1929. In 1927, only one public pawnshop was present in the Shimoya, Asakusa, Koishikawa, and Kyobashi wards, whereas the Fukagawa and Honjo wards had two each. Refer to Figure 2 for the location of each ward. Sources: Tokyo City Social Welfare Bureau (1928–1935); Statistical Division of the Commerce and Industry Minister (various years).

Figure 3 plots the time-series data on the loan amounts provided by public pawnshops in various wards from 1927 to 1935, highlighting diverse trends in different wards. Specifically, the loan amounts exhibit an upward trajectory in Fukagawa and Shimoya wards during this period. In contrast, those in Koishikawa and Kyobashi wards remain relatively flat without substantial fluctuations. In the case of Honjo and Asakusa wards, the loan amounts grow throughout the 1920s but slightly decrease in the 1930s. Shiba ward, where a public pawnshop is established for the first time in 1930, experiences a rapid loan increase for the first 2 years. However, this growth stagnates after that.[9] The observed disparities among the wards do not depend on whether the public pawnshops are established earlier or on the loan amounts recorded in 1927, the initial year of the study period.

Figures 2 and 3 present two significant observations regarding public pawnshops. First, they are more commonly established in areas characterized by relatively low incomes. Second, the growth of public pawnshops displays varying trends across wards. These insights aid the empirical analysis of the health impacts of public pawnshops using ward-level panel data. As ward fixed effects control for variations in initial endowments, I leverage heterogeneity in the within-variation of loan amounts to identify the within-estimator.

---

[9] Figure A.1 in Appendices shows the loan amounts of private pawnshops in each ward from 1927 to 1935, indicating no clear link to those of public pawnshops. Figure A.2, which illustrates the number of private pawnshops in each ward, also supports no apparent relationship to the increase in public pawnshops. These comparisons suggest the differences in customers and purpose of use between public and private pawnshops. In wards without public pawnshops, the number and amount of private pawn loans remained relatively constant.

In Tokyo, pawnshops were a popular financial resource for low-income households requiring money in the 1920s and 1930s.[10] A 1933 survey of 15,634 poor households in Tokyo reveals that 73% (4,644 of 6,355 borrowings) of all loans are obtained through pawnshops. Of the 4,644 pawn loans, 99% (4,605) use pawning clothes as collateral (Tokyo Prefecture Department of Academic Affairs, 1935). This crucial role of clothes as collateral is further corroborated by other surveys, which indicate that clothes constitute 76%–81% of all items pawned in Tokyo's public pawnshops in 1923 (Social Welfare Bureau, 1926c; Tokyo Institute for Municipal Research, 1926).

The 1933 survey reveals that 85% of the total loan amounts are used to cover living expenses and approximately 9% are allocated to medical expenses.[11] The use of pawn loans for these purposes suggests that such loans potentially contribute to health improvements in low-income households. As described later, by helping them maintain better hygiene and nutrition, pawn loans are instrumental in reducing the mortality risk, especially among vulnerable infants and fetuses.

[10] Ioku and Shizume (2014) focusing on one pawnshop in prewar Tokyo emphasize the importance of pawnshops as a financial institution for common people.

[11] Two surveys are conducted in prewar Osaka City to investigate how low-income households use loans from pawnshops. The first is conducted between 1923 and 1924. It reveals that households use 50.94% and 7.16% of the amount borrowed from public pawnshops for living and medical expenses, respectively (Tokyo Institute for Municipal Research, 1926). Another survey in 1941 reports that 58.90% of loans from private and public pawnshops is used for meeting living costs and 35.62% for medical costs (Osaka City Social Welfare Bureau, 1942).

### 2.3. Differences between Public and Private Pawnshops

The pawnbroking systems of public and private pawnshops are mostly similar. People used both types of institutions in the same manner to secure loans, using their pawned items as collateral. However, there were some important differences between public and private pawnshops, driven by the purpose of public pawnshops to support economically disadvantaged people.

A major distinction between public and private pawnshops was the interest rates charged on loans. The Public Pawnshop Law mandated lower interest rates for public institutions than for their private counterparts. While private pawnshops charged a monthly interest rate of 2.5%–4.0% according to the Pawnshop Control Law, public pawnshops only charged 1.25%.[12] This lower rate enabled borrowers to access loans from public pawnshops at less than half the interest rate of private pawnshops. This reduction in the financial burden was particularly beneficial for low-income households, as smaller loans had higher interest rates in private pawnshops.

Another difference between public and private pawnshops was the ceiling on loan amounts. The Public Pawnshop Law imposed maximum loan amounts of 10 yen per loan and 50 yen per household to prevent individuals other than the low-income class from exploiting public pawnshops.[13] This restriction was supported by private pawnshop owners, who were concerned about competition from public institutions and wanted to limit their scale of operation. The Tokyo City Office (1926) indicate that this upper limit effectively achieved its purpose. Loans exceeding 10 yen accounted for approximately 25% of all the loans provided by private pawnshops (two-thirds of the total value), suggesting that some people secured relatively large loans from these institutions.[14] Therefore, the limitation on loan amounts allowed public pawnshops to exclude middle-income borrowers, such as merchants who required larger loans, and focus on lending money to low-income individuals at low interest rates.

---

[12] As private pawnshops do not change their interest rates depending on those of other financial institutions or the economy, the rates generally remain constant (Bank of Japan, 1913). Regarding deposit interest, annual interest rates in December 1930 are 4.84% and 4.20% on six-month fixed deposits in Tokyo's banks and postal savings, respectively. In addition, the yield of government bonds is 5.50% (Bank of Japan, 1986).

[13] According to the Tokyo Chamber of Commerce and Industry (1935), mean wage per day of workers is approximately 2.07 yen in Tokyo during 1927–1935, and day workers' wage is 1.23 yen. Additionally, interest rates and upper limit amounts of public pawnshops vary based on regions and managing organizations (Social Welfare Bureau, 1926a). Tokyo changed the maximum loan amount to 20 yen per loan and 100 yen per household in 1930 (Tokyo Prefecture Department of Academic Affairs, 1935).

Overall, although public and private pawnshops shared many similarities in their systems, disparities in interest rates and loan ceilings shaped their roles in providing financial services. Public pawnshops can effectively target economically struggling classes and offer much-needed financial support. As these financially vulnerable populations often faced poor health conditions, public pawnshops, despite their small numbers, may have contributed more to health improvements than private pawnshops.

---

[14] Tokyo City Office (1926) reports the number and amount of loans provided by private pawnshops in 1924 based on groups of amounts. Loans of 10–20, 20–50, 50–100, and more than 100 yen constituted 15.13%, 6.86%, 1.83%, and 0.50% of the total number of loans, respectively. Regarding value, they represent 23.37%, 21.56%, 14.98%, and 8.57% of the total, respectively.

## 3. Mechanisms

This section discusses how pawn loans decreased infant and fetal death rates. Poverty-related health problems stem from factors such as limited access to medical care, malnutrition due to food insecurity, and poor hygiene. However, medical treatments were less influential in health improvement than today, as effective medicines such as antibiotics were not developed in the prewar period. Hence, the primary benefits of pawn loans were enhanced nutritional intake and the maintenance of hygiene. The importance of these factors is supported by a contemporary survey of lowest-income households, which reports strong relationships among poor hygiene, maternal malnutrition, poor family health, and infant and fetal deaths (Tokyo City Office, 1937).

Although securing money through access to loans contributed to various health improvements, the impact on pneumonia and diarrhea seemed particularly important because these diseases were the two major causes of infant mortality during the interwar period in Tokyo. According to Tokyo City (1931), 18.84% and 17.90% infant deaths are caused by pneumonia and diarrhea, respectively. Moreover, maternal infection and nutritional deficiencies affect fetal health.[15] Thus, pawnshop loans led to reductions in infant and fetal deaths by mitigating the risks of these health issues.

Inadequate nutrition causes immunological deterioration and increases the risk of contracting pneumonia and other infectious diseases. Moreover, maternal malnutrition causes infant diarrhea mortality by reducing the quantity and quality of breast milk. In interwar Tokyo, mothers who had trouble breast feeding used breast milk substitutes, such as cow's milk and condensed milk.[16] A 1934 survey of 16,827 protection-requiring households reports that 1,138 of 5,153 infants (22.08%) are fed substitutes (Tokyo City Social Welfare Bureau, 1935b).[17] However, bottle feeding cause more infant mortality, especially deaths from diarrhea, potentially because of indigestion and poor nutrition, than breast feeding (Santaya, 1941).[18] Another survey on infant deaths conducted in the

---

[15] Maternal infection can cause fetal death due to high fever, respiratory distress, and other systemic reactions (Goldenberg et al., 2010).

[16] The price of milk is approximately 0.08–0.11 yen per 200 milliliters, while that of a can of condensed milk is approximately 0.32–0.48 yen (Tokyo City, 1931; 1937).

[17] Although Tokyo offered milk rationing programs for infants in poor families who have difficulty breast feeding, these are insufficient owing to lack of funds (Tokyo City Social Welfare Bureau, 1926; 1929).

[18] No households spend more money to provide additional nourishment for their bottle-fed infants in the 1934 survey (Tokyo City Social Welfare Bureau, 1935b). Additionally, as milk is perishable, the risk of

Koishikawa, Shimoya, and Honjo wards between 1924 and 1925 reported consistent findings. Among breast-fed infants, the leading cause of death was congenital debility, whereas among bottle-fed infants, this was diarrhea, followed by pneumonia (Tokyo City Social Welfare Bureau, 1926b).

During the late 1920s and 1930s, poorly fed children (*kesshoku jido*) attracted significant public attention as a pressing social issue. They could not bring a boxed lunch (*bento*) to school or, in some cases, even eat breakfast because of poverty, underscoring that economically disadvantaged families were undernourished at that time. [19] Furthermore, a 1931 living survey of 21,666 low-income households in Tokyo suggests that the monthly average food cost is 18.97 yen per household and 4.69 yen per capita, approximately 54.51% of the total household income (Tokyo City Office, 1932). This high proportion suggests that the quantity of daily diet largely depends on daily income. In this context, loans provided by public pawnshops helped with food purchases and

---

bottle feeding is higher in low-income households without a refrigerator (including an old-fashioned refrigerator) than in high-income households (Santaya, 1941).

[19] Although the Tohoku region in northeastern Japan experiences severe poverty, Tokyo is no exception in this issue. A survey by the Tokyo City Social Welfare Bureau (1932c) finds that over half of elementary schools (111 of 203) have poorly fed children, totaling 2,847 students.

nutritional intake under income shocks through consumption smoothing, thereby contributing to reduced infant mortality and fetal death rates.

Furthermore, public pawnshops could have played a crucial role in enhancing the health of low-income households by facilitating access to essential hygiene goods and services. One potential pathway was the purchase of daily necessities such as soap, which was a cost-effective way to prevent many respiratory and digestive infectious diseases. According to a 1921 survey, even poor households buy soaps to keep them clean (Bureau of Social Affairs, 1922).[20]

Another possibility for loan use for better hygiene was paying bathing fees in public bathhouses (*sento*). Bathing is an effective preventive measure against illness, as it not only cleans the body but also boosts the immune system by increasing the body temperature. Nonetheless, during the economic recession of the late 1920s and 1930s, the cost of using public bathhouses (0.05 yen per adult in the private sector) is a considerable financial burden for many low-income individuals.[21] Consequently, they cannot bathe frequently. This issue is serious enough to raise concerns of the city government from a hygienic standpoint (Kawabata, 2016). Given this historical situation, low-interest loans from public pawnshops helped low-income individuals pay bathing fees and improve their health outcomes.

---

[20] Soap is not a luxury good and widely prevalent in the 1920s; the price of a piece of *Kao sekken*, the most popular soap, is approximately 0.15 yen, which further reduced to about 0.10 yen with the launch of a new product in 1931 (Kao Museum and Material Room, 2012; Tokyo City, 1931; 1937).

In a related channel, covering fuel (firewood and charcoal) costs for heating in winter reduced infant mortality due to respiratory diseases such as pneumonia.[22] Low air temperature causes immune compromise through a drop in body temperature, in addition to viral activation. Maintaining the body temperature required many calories, which was difficult for low-income families with inadequate nutritional intake. Moreover, poor residential environments, attributable to poverty, exacerbated the problem of low temperatures. This hypothetical channel is consistent with the seasonality of public

---

[21] In Tokyo, campaigns promoting the reduction of bathing fees begin in 1930. Consequently, city-funded public bathhouses reduced the fee from 0.04 yen to 0.03 yen in 1932. Additionally, some private-sector public bathhouses distributed free or discounted coupons as charities. For more information, refer to Kawabata (2016).

[22] According to a 1934 survey of 186 low-income households in the new city area of Tokyo, the mean utility cost is 1.88 yen, constituting 7.63% of the total household expenditure (Tokyo City Social Welfare Bureau, 1935a).

pawnshop loans in 1927–1935 and infant mortality in 1926, both of which are higher in winter than summer.[23]

Furthermore, public pawn loans were used to cover pregnancy and childbirth expenses. Based on a survey of 460 low-income households who experienced childbirth between September 1926 and March 1927, the Tokyo City Social Welfare Bureau reports that the average delivery cost is 18.30 yen (Tokyo Institute for Municipal Research, 1928). Despite the anticipated need for months of maternity-related costs, 23.48% of these households resort to borrowing, including pawn loans, to meet their expenses. In addition, although Tokyo has three maternity hospitals designated for low-income households during this period, their capacity is inadequate to accommodate the increasing demand.[24] Consequently, many pregnant women with economic difficulties give birth at home, even if they require hospitalization (Tokyo City Social Welfare Bureau, 1930; 1936).[25] If public pawn loans were allocated toward maternity expenses, such as hospitalization costs, fetal deaths could be averted.[26]

[23] Although the monthly data on infant mortality is unavailable for the study period, Figure A.3 in Appendices shows the monthly mean of loan amounts provided by public pawnshops during 1927–1935 and the monthly number of infant deaths in 1926.

[24] These maternity hospitals were established in Kyobashi in 1924, Asakusa in 1925 (relocated to Shimoya in 1933), and Fukagawa in 1927.

## 4. Methods

### 4.1. Data

For the regression analyses, I constructed a ward-level balanced panel dataset from 15 wards during 1927–1935 using various official documents. The data sources for the amount and number of public pawnshop loans are annual reports from the Tokyo City Social Welfare Bureau (*Tokyoshi Shakaikyoku Nempo*). As these reports recorded the monthly data of every public pawnshop, they were aggregated at the ward-year level.[27] To retain observations with zero values, this study applies an inverse hyperbolic sine transformation instead of taking the natural logarithm of the loan variables.[28] Data on private pawnshops are described in the Statistical Yearbooks of Tokyo Prefecture (*Tokyofu Tokeisho*), published by the prefectural government.[29]

[25] From October 1929, Asakusa (Shimoya) and Fukagawa maternity hospitals dispatched midwives to mothers who required hospitalization but could not afford it owing to dire poverty throughout Tokyo. However, this service was also limited compared with the demand. Even the highest number of assisted births was 743 in 1931, which was only 1.39% of total births, and the lowest was 96 (0.22%) in 1934.

[26] Additionally, a 1941 survey in Osaka lends partial support to this possibility, as it reports that pawn loans are employed to cover childbirth expenses (Osaka City Social Welfare Bureau, 1942).

The health outcomes assessed were infant and fetal death rates. The infant mortality rate is the number of infant deaths per 1,000 live births, whereas the fetal death rate is the number of stillbirths per 1,000 births.[30] These death rates are ideal measurements for this study. When the breadwinner of a family died, the bereaved family might borrow money from pawnshops to cover the loss of income. Focusing on the health status of infants and fetuses allowed to avoid reverse causality. Moreover, as they were more sensitive and vulnerable to poor sanitary and nutritional conditions than adults, the effects of health improvement efforts through pawnshop loans could be captured. Furthermore, reducing early life mortality risk is important for assessing the value of public investment because of the expected long-term economic activity in the future. Data on infant and fetal deaths were collected from the Statistical Yearbooks of Tokyo City (*Tokyoshi Tokei Nempyo*), published by the Tokyo City Office. These yearbooks covered all infant and fetal deaths, and the information was compiled using official registration-based statistics. Thus, the study's statistical inference was unaffected by sample selection bias.

[27] The monthly data on public pawn loans are reported for each fiscal year. Thus, we estimated figures at the calendar year-level for the first and final years (1927 and 1935) based on the figures from April to December and January to March, respectively.

[28] For a random variable $x$, this transformation creates $\mathrm{arsinh}(x) = \ln\left(x + \sqrt{x^2 + 1}\right)$. The marginal effect of $\mathrm{arsinh}(x)$ approximately equals to that of $\ln(x)$ for a sufficiently large $x$.

[29] The loan amounts of public and private pawnshops used in quantitative analyses are the real values adjusted by the wholesale price index based on December 1929.

[30] In prewar Japan, stillbirth was defined as the childbirth of a dead fetus after the fourth month of pregnancy.

The control variables include the coverage of social workers, proportion of taxpayers, coverage of medical doctors, and coverage of modern water taps, obtained from the Statistical Yearbooks of Tokyo City. These covariates represent levels of social welfare, income, popularization of medical treatments, and public health.

Table 1: Summary Statistics

| | Mean | SD | Min | Max | Obs. |
|---|---|---|---|---|---|
| Dependent Variables (‰) | | | | | |
| Infant mortality rate | 105.30 | 31.87 | 44.72 | 196.72 | 135 |
| Fetal death rate | 59.82 | 14.77 | 27.65 | 109.36 | 135 |
| Pneumonia death rate | 1.30 | 0.33 | 0.63 | 2.13 | 45 |
| Diarrhea death rate | 0.66 | 0.28 | 0.14 | 1.26 | 45 |
| Congenital debility death rate | 0.87 | 0.31 | 0.34 | 1.83 | 45 |
| Female mortality from infection | 0.39 | 0.10 | 0.12 | 0.66 | 135 |
| Key Independent Variables | | | | | |
| Pawn loan amount (yen) | | | | | |
| Public | 58,939.70 | 85,017.87 | 0.00 | 323,745.25 | 135 |
| Private | 1,483,077.42 | 687,504.23 | 469,815.92 | 3,444,126.99 | 135 |
| Number of pawn loans | | | | | |
| Public | 12,634.37 | 19,100.20 | 0.00 | 84,004.00 | 135 |
| Private | 223,451.41 | 121,851.84 | 53,368.00 | 607,817.00 | 135 |
| Control Variables (%) | | | | | |
| Social worker rate | 0.03 | 0.02 | 0.00 | 0.07 | 135 |
| Taxpayer rate | 3.92 | 1.44 | 1.32 | 6.69 | 135 |
| Doctor rate | 0.22 | 0.13 | 0.06 | 0.64 | 135 |
| Water tap rate | 36.02 | 18.03 | 4.68 | 70.68 | 135 |
| Interacted Variables | | | | | |
| Poorest wards | 0.27 | 0.46 | 0.00 | 1.00 | 15 |
| Public loan amount in 1927 | 25,700.62 | 37,486.35 | 0.00 | 106,928.70 | 15 |
| Poverty ratio (%) | 5.98 | 5.19 | 0.05 | 25.04 | 75 |

Notes: The infant mortality rate is defined as the number of infant deaths per 1,000 live births, while the fetal death rate is defined as the number of stillbirths per 1,000 births (sum of live births and fetal deaths). The pneumonia, diarrhea, and congenital death rates are defined as the number of deaths due to respective causes per 1,000 people. Female mortality from infection is defined as the number of female deaths due to infectious diseases per 1,000 people. The pawn loan amount represents the total amount (in yen) of loans from public or private pawnshops, and the number of pawn loans indicates the total number of loans from either type of pawnshop. The loan amounts are the real values adjusted by the wholesale price index based on December 1929. The social worker, taxpayer, doctor, and water tap rates are defined as the number of social workers, taxpayers, doctors, and modern water taps per 100 population, respectively. The dummy variable for the poorest wards is set to one if the ward is Shimoya, Asakusa, Honjo, or Fukagawa. The public loan amount in 1927 refers to the total amount loaned by public pawnshops in each ward during 1927. The poverty ratio denotes the percentage of protection-requiring people in relation to the total population. Sources: Tokyo City (1929–1937); Tokyo City Office (1933a); Tokyo City Social Welfare Bureau (1928–1935); Tokyo City Social Welfare Bureau (1930b–1932b); Tokyo Prefecture (1929–1937).

Table 1 presents the summary statistics of the variables, emphasizing the remarkable difference in the amounts loaned by public and private pawnshops. The mean value of the amount loaned by public pawnshops is 58,934 yen or 126,299 yen, excluding wards without public pawnshops. These amounts are approximately 3.97% and 8.52% of the total loans made by private pawnshops. This disparity persists when measured using the number of loans. Despite the decline in private pawnshops following the Great Kanto Earthquake and expansion of public pawnshops in the late 1920s, the scale of public institutions' operations remained far below that of their private counterparts.

### 4.2. Identification Strategy

Considering the panel data structure, I employ a fixed-effects model to investigate the relationship between pawn loans and health outcomes. The baseline specification is as follows:

$$y_{it} = \alpha + \beta PubLoan_{it} + \gamma PriLoan_{it} + \boldsymbol{x}'_{it}\boldsymbol{\delta} + \nu_i + \mu_t + t\theta_i + \varepsilon_{it},$$

where $y_{it}$ is the natural logarithm of infant or fetal death rate. $PubLoan_{it}$ and $PriLoan_{it}$ represent the total loans from public (inverse hyperbolic sine) and private pawnshops (natural logarithm), respectively. $\boldsymbol{x}_{it}$ is a vector of control variables in ward $i$ and year $t$, and $\nu_i$ and $\mu_t$ are the ward and year fixed effects, respectively. The $t\theta_i$ denotes ward-specific linear time trends, and $\varepsilon_{it}$ is a random error term.

Figure 2 shows that pawnshops are more established in low-income areas, where the population generally experiences poor health. This raises a concern that the independent variables representing pawn loans correlate with households' unobservable living standards in the error term. Although identifying the exact causal effects is challenging, this regression model addresses the potential omitted variable bias in the following three ways to facilitate a deeper understanding of the impact of pawnshops on health outcomes:

First, I include observable characteristics that captured the wealth levels in each ward. These include social worker coverage, taxpayer proportions, medical doctor shares, and modern water tap coverage. These ward-year-level variables help control for potential correlations between the pawning variables and error term. Second, I incorporate ward fixed effects ($\nu_i$) to control for all unobservable time-constant factors, such as the geographical characteristics and preferences of local ward offices. Furthermore, I include year fixed effects ($\mu_t$) to account for unobservable macroeconomic shocks, such as the depression around 1930.[31] Finally, I consider ward-specific time trends ($t\theta_i$) to address heterogeneity in the trends of infant and fetal deaths.

Standard errors are clustered at the ward level to allow for potential correlations among errors in each ward and heteroscedasticity across wards. All regressions are weighted using the mean values of the denominators of the dependent variable.

## 5. Results

### 5.1. Main Results

Table 2 reports the estimation results of the loan amounts from public and private pawnshops. Columns (1)–(4) show the results for the infant mortality rate, while columns (5)–(8) display those for the fetal death rate. Columns (1) and (5) incorporate the fixed effects, time trends, and coverage of social workers as a control variable. Columns (2) and (6) further assimilate the proportion of taxpayers, while columns (3) and (7) include

[31] Several studies on Japanese history reveal that the economy recovered quickly because the impact of economic depression is relatively small at that time (Miyamoto 2008, pp. 56–57).

Table 2: Relationships between Pawn Loan Amounts and Health Outcomes

| | Infant mortality rate | | | | Fetal death rate | | | |
|---|---|---|---|---|---|---|---|---|
| | (1) | (2) | (3) | (4) | (5) | (6) | (7) | (8) |
| Public loans | -0.024** | -0.023*** | -0.023*** | -0.022*** | -0.033*** | -0.033*** | -0.032*** | -0.029*** |
| | (0.008) | (0.008) | (0.008) | (0.007) | (0.006) | (0.006) | (0.006) | (0.006) |
| Private loans | -0.015 | -0.036 | -0.036 | -0.043 | 0.239 | 0.229 | 0.226 | 0.211 |
| | (0.155) | (0.158) | (0.159) | (0.164) | (0.186) | (0.179) | (0.177) | (0.185) |
| Social worker | Yes | Yes | Yes | Yes | Yes | Yes | Yes | Yes |
| Taxpayer | No | Yes | Yes | Yes | No | Yes | Yes | Yes |
| Doctor | No | No | Yes | Yes | No | No | Yes | Yes |
| Water tap | No | No | No | Yes | No | No | No | Yes |
| Ward FE | Yes | Yes | Yes | Yes | Yes | Yes | Yes | Yes |
| Year FE | Yes | Yes | Yes | Yes | Yes | Yes | Yes | Yes |
| Time trend | Yes | Yes | Yes | Yes | Yes | Yes | Yes | Yes |
| Observations | 135 | 135 | 135 | 135 | 135 | 135 | 135 | 135 |

***, **, and * represent statistical significance at the 1%, 5%, and 10% levels, respectively.
Notes: All regressions in columns (1)–(4) and (5)–(8) are weighted by the number of live births and that of births, respectively. The dependent variables and private pawn loans are logarithms, while the inverse hyperbolic sine transformation is applied to public pawn loans. The loan amounts are the real values. Standard errors clustered at the ward level are in parentheses.

the share of doctors. Finally, I add the modern water tap coverage in columns (4) and (8), which present the results of baseline specification.

Column (1) shows that the coefficient of the amount loaned by public pawnshops is negative and statistically significant. This result remains unchanged when additional control variables are introduced in columns (2)–(4). These negative relationships suggest that loans from public pawnshops improve health, leading to a decline in infant mortality rates. In addition, the magnitude of the baseline result in column (4) indicates that a 1% increase in public pawn loan amounts is associated with a 0.022% decrease in infant mortality rate. Given the 167.87% increase in the mean loan amount from 385,509 yen in 1927 to 1,032,671 yen in 1935, a simple calculation implies that public pawn loans are related to a 3.69% reduction in the infant mortality rate during this period. This finding emphasizes the importance of microfinancial institutions in promoting positive health outcomes.

Private pawnshops do not exhibit a significant relationship with infant mortality rates. Unlike public pawn loans, the coefficients of private pawn loans are negative but statistically nonsignificant regardless of the control variables. Although private pawnshops maintained their position as the primary financial institutions for low-income households alongside public institutions, the study's results suggest that they are not associated with improvements in health conditions during the late 1920s and 1930s.

These contrasting results for the public and private pawnshops have important implications. Despite the functional similarities and popularity of private pawnshops, they are less influential in improving health than their public counterparts. Thus, minor institutional differences likely enabled public pawnshops, despite their smaller scale, to improve health outcomes of low-income people. These findings imply that small changes favoring low-income households, such as interest rate reductions and loan caps, can substantially and positively impact their well-being through easing liquidity constraints.

Columns (5)–(8) present the results for the fetal death rate, which are similar to those for the infant mortality rate. The coefficient of public pawn loans is significantly negative and remains stable when various control variables are included. By contrast, private pawn loans show no significant association with fetal death across all specifications. These findings highlight the crucial role of public institutions in improving health, which is consistent with their impact on infant mortality. The baseline result in column (8) indicates that a 1% increase in the amount loaned by public pawnshops is associated with a 0.029% decrease in the fetal death rate. Consequently, the rise in public pawn loans contributes to a 4.87% reduction in the fetal death rate in Tokyo during 1927–1935.

### 5.2. Additional Analyses

Specific applications of pawn loans are unobservable.[32] Nevertheless, this study confirms the validity of potential channels through which loans decrease infant and fetal deaths by exploring the relationships between public pawn loans and death rates from diarrhea and pneumonia. As ward-level data on the causes of infant deaths are unavailable, I used data on pneumonia and diarrhea death rates for all ages in every ward between 1928 and 1930, sourced from the Tokyo City Social Welfare Bureau (1930b–1932b).[33] Additionally, the death rate due to congenital debility, another major cause of infant death at the time, was used as a placebo test. This cause was less affected by potential mechanisms, such as improvements in nutrition and hygiene; thus, its death rate should not significantly correlate with the increase in public pawn loans.

[32] The unavailability of direct data on the use of loans is a common challenge, regardless of whether these are historical studies (Bhutta, 2014; McLaughlin, 2022).

Table 3 presents the results for the cause-specific death rate. The outcome variables are the logarithm of deaths per 1,000 people due to pneumonia in columns (1) and (2), diarrhea in columns (3) and (4), and congenital debility in columns (5) and (6). All estimation equations incorporate ward and year fixed effects.[34]

---

[33] These surveys on mortality and fetal deaths are conducted to provide social welfare services. In 1929, infant deaths accounted for 38.33% and 51.39% of deaths due to pneumonia and diarrhea for all ages, respectively (Tokyo City, 1931). These substantial proportions indicate that a reduction in pneumonia and diarrhea cases cause a decline in infant mortality.

[34] These regression analyses do not include the ward-specific linear time trend because of their small sample size.

Table 3: Relationships between Pawn Loans and Cause-Specific Death Rates

| | Pneumonia | | Diarrhea | | Congenital debility | |
|---|---|---|---|---|---|---|
| | (1) | (2) | (3) | (4) | (5) | (6) |
| Public loans | -0.013*** | -0.016*** | -0.019* | -0.009 | -0.008 | -0.003 |
| | (0.003) | (0.003) | (0.010) | (0.008) | (0.009) | (0.008) |
| Private loans | -0.055 | -0.061 | 0.420 | -0.135 | 0.220 | -0.134 |
| | (0.166) | (0.250) | (0.523) | (0.516) | (0.316) | (0.323) |
| All control variables | No | Yes | No | Yes | No | Yes |
| Ward FE | Yes | Yes | Yes | Yes | Yes | Yes |
| Year FE | Yes | Yes | Yes | Yes | Yes | Yes |
| Time trend | No | No | No | No | No | No |
| Observations | 45 | 45 | 45 | 45 | 45 | 45 |

***, **, and * represent statistical significance at the 1%, 5%, and 10% levels, respectively.
Notes: All regressions are weighted by population. The dependent variables and private pawn loans are logarithms, while the inverse hyperbolic sine transformation is applied to public pawn loans. All control variables are social worker, taxpayer, doctor, and water tap rates. Ward-specific linear time trends are not included in the regressions due to the small sample size. Standard errors clustered at the ward level are in parentheses.

The coefficients of public pawn loans are significantly negative for the pneumonia death rate regardless of whether the control variables are included. Column (2) indicates that a 1% increase in the amount loaned by public pawnshops corresponds to a 0.016% decline in the pneumonia mortality rate. This negative correlation between public pawn loans and deaths due to pneumonia is consistent with this study's hypotheses and primary findings, suggesting the contribution of public pawnshops in improving infant health through nutritional intake, better hygiene, and access to heating.

Furthermore, the results for diarrheal death rate are negative. While the estimated coefficient is statistically significant in column (3), it becomes lower and nonsignificant in column (4), where all control variables are added. Therefore, public pawnshops may have contributed to the reduction in infant mortality by mitigating the risk of pneumonia rather than diarrhea. However, another possibility exists that this change in significance is attributed to the small sample size, comprising only 3 years of data for the 15 wards.

The results in columns (5) and (6) further support the validity of the potential channels. As expected, the death rate due to congenital debility is unassociated with pawn loans. In contrast to public pawn loans, those from private pawnshops show no significant relationship with either the pneumonia or diarrhea death rates in columns (1)–(4). These results correspond to the main findings and highlight the role of public pawnshops in enhancing well-being.

Maternal infection may have been a factor in the relationship between public pawn loans and improvements in fetal deaths. To examine whether public pawnshops reduced the fetal death rate by helping low-income pregnant women avoid infection, I investigated the relationship between the loan amounts of public pawnshops and female mortality due to infectious diseases using regression analyses of ward-level data from 1925 to 1937. The dependent variable is the logarithm of the number of female deaths due to infectious

Table 4: Relationships between Pawn Loans and Female Mortality from Infection

| | Female death rate due to infectious diseases | | | |
|---|---|---|---|---|
| | (1) | (2) | (3) | (4) |
| Public loans | -0.012* | -0.012* | -0.012* | -0.011 |
| | (0.007) | (0.006) | (0.006) | (0.007) |
| Private loans | 0.024 | 0.029 | 0.029 | 0.025 |
| | (0.101) | (0.091) | (0.089) | (0.092) |
| Social worker | Yes | Yes | Yes | Yes |
| Taxpayer | No | Yes | Yes | Yes |
| Doctor | No | No | Yes | Yes |
| Water tap | No | No | No | Yes |
| Ward FE | Yes | Yes | Yes | Yes |
| Year FE | Yes | Yes | Yes | Yes |
| Time trend | Yes | Yes | Yes | Yes |
| Observations | 135 | 135 | 135 | 135 |

***, **, and * represent statistical significance at the 1%, 5%, and 10% levels, respectively.
Notes: All regressions are weighted by population. The dependent variables and private pawn loans are logarithms, while the inverse hyperbolic sine transformation is applied to public pawn loans. The loan amounts are the real values. Standard errors clustered at the ward level are in parentheses.

diseases per 1,000 individuals.[35] The estimating equations, other than the outcomes, are similar to those in Table 2, including the control variables, ward and year fixed effects, and ward-specific linear time trends.

Table 4 shows the results, in which the estimated coefficients of public pawn loans are negative across all specifications. These negative relationships suggest that low-interest loans provided by public pawnshops may reduce the risk of maternal infection through improved nutrition and hygiene, leading to a decline in fetal death rate. However, the statistically nonsignificant result in column (4), incorporating all control variables, implies that this potential pathway is nonrobust compared with the channel through contracting pneumonia. Similar to the estimation results for the other health outcomes, private pawnshops have no significant correlation.

[35] I sourced the number of female deaths from infectious diseases from the Statistical Yearbooks of Tokyo City.

### 5.3. Robustness checks

Regression analyses are performed to examine the robustness of the primary findings. First, I use the number of pawn loans as the key independent variable instead of the loan amount. This alternative measurement verifies whether the estimation results remain consistent regardless of how pawn loans are measured. Table 5 shows the relationship between the number of pawn loans and health outcomes. The dependent variables are the infant mortality rate in columns (1)–(3) and fetal death rate in columns (4)–(6). The independent variables of interest are the number of loans from public pawnshops in columns (1) and (4), private pawnshops in columns (2) and (5), and both in columns (3) and (6). Consistent with the baseline specifications used in columns (4) and (8) of Table 2, all specifications include all control variables, ward- and year-fixed effects, and ward-specific time trends.

Table 5: Relationships between the Number of Pawn Loans and Health Outcomes

| | Infant mortality rate | | | Fetal death rate | | |
|---|---|---|---|---|---|---|
| | (1) | (2) | (3) | (4) | (5) | (6) |
| Number of public loans | -0.028*** | | -0.028*** | -0.035*** | | -0.037*** |
| | (0.008) | | (0.009) | (0.009) | | (0.009) |
| Number of private loans | | -0.030 | -0.000 | | 0.140 | 0.179 |
| | | (0.110) | (0.103) | | (0.125) | (0.113) |
| All control variables | Yes | Yes | Yes | Yes | Yes | Yes |
| Ward FE | Yes | Yes | Yes | Yes | Yes | Yes |
| Year FE | Yes | Yes | Yes | Yes | Yes | Yes |
| Time trend | Yes | Yes | Yes | Yes | Yes | Yes |
| Observations | 135 | 135 | 135 | 135 | 135 | 135 |

***, **, and * represent statistical significance at the 1%, 5%, and 10% levels, respectively.
Notes: All regressions in columns (1)–(3) and (4)–(6) are weighted by the number of live births and that of births, respectively. The dependent variables and private pawn loans are logarithms, while the inverse hyperbolic sine transformation is applied to public pawn loans. All control variables are social worker, taxpayer, doctor, and water tap rates. Standard errors clustered at the ward level are in parentheses.

Consequently, these analyses yield findings similar to those obtained using the loan amounts. Columns (1), (3), (4), and (6) suggest that the number of public pawn loans is significantly and negatively associated with infant and fetal death rates. By contrast, the coefficients of the number of private pawn loans are nonsignificant for either outcome in columns (2), (3), (5), and (6). These findings correspond with the main results and demonstrate their robustness.

The statistically significant results in columns (3) and (6) suggest that a 1% increase in the number of public pawn loans is associated with 0.028% and 0.037% reductions in infant and fetal death rates, respectively. These relationships imply that the increase in the number of public pawn loans between 1927 and 1935 contribute to a 4.30% decline in infant mortality and 5.68% decline in fetal death rates. Comparing these results with the main findings, the elasticity of the number of loans is greater than that of the loan amount. This heightened sensitivity indicates that the frequency of taking loans is more effective than their amount in maintaining better health.

Although the proportion of small loans among the total private pawn loans is larger in volume than in value, the estimates of private pawnshops indicate that the correlation between private pawn loans and health outcomes remains weak even when measured based on quantity. These results emphasize the contribution of public pawnshops to improvements in population health, relative to private institutions. It seems that low-income households use public rather than private pawn loans to enhance their hygiene and nutritional status.

Second, I employ interaction models to investigate the potential heterogeneity in the relationship between public pawn loans and health outcomes. While low-income households primarily use public pawnshops, middle- and high-income individuals do not rely on pawnshops as financial resources to cope with their health concerns. Thus, the relationship between public pawn loans and health improvement is more evident in

Table 6: Heterogeneity of Relationships between Pawn Loans and Health Outcomes

| | Infant mortality rate | | | | Fetal death rate | | | |
|---|---|---|---|---|---|---|---|---|
| | (1) | (2) | (3) | (4) | (5) | (6) | (7) | (8) |
| Public loans | -0.020** | -0.019* | -0.035*** | -0.019 | -0.028*** | -0.028*** | -0.033** | -0.030*** |
| | (0.009) | (0.009) | (0.009) | (0.015) | (0.007) | (0.008) | (0.011) | (0.010) |
| Public loans × | | | | | | | | |
| Poorest wards | 0.107 | | | | 0.153 | | | |
| | (0.119) | | | | (0.135) | | | |
| 1927 Public loans | | 0.014 | | | | 0.011 | | |
| | | (0.009) | | | | (0.010) | | |
| Social worker | | | 0.839 | | | | 0.212 | |
| | | | (0.824) | | | | (0.674) | |
| Poverty ratio | | | | 0.000 | | | | 0.000 |
| | | | | (0.001) | | | | (0.001) |
| Private loans | -0.047 | -0.098 | -0.062 | -0.036 | 0.203 | 0.166 | 0.206 | 0.229 |
| | (0.162) | (0.172) | (0.177) | (0.199) | (0.169) | (0.183) | (0.185) | (0.178) |
| All control variables | Yes | Yes | Yes | Yes | Yes | Yes | Yes | Yes |
| Ward FE | Yes | Yes | Yes | Yes | Yes | Yes | Yes | Yes |
| Year FE | Yes | Yes | Yes | Yes | Yes | Yes | Yes | Yes |
| Time trend | Yes | Yes | Yes | Yes | Yes | Yes | Yes | Yes |
| Observations | 135 | 135 | 135 | 75 | 135 | 135 | 135 | 75 |

***, **, and * represent statistical significance at the 1%, 5%, and 10% levels, respectively.
Notes: All regressions in columns (1)–(4) and (5)–(8) are weighted by the number of live births and that of births, respectively. The dependent variables and private pawn loans are logarithms, while the inverse hyperbolic sine transformation is applied to public pawn loans. The loan amounts are the real values. All control variables are social worker, taxpayer, doctor, and water tap rates. Columns (4) and (8) cover only years 1929, 1931, 1932, 1933, and 1935 due to a lack of data. Standard errors clustered at the ward level are in parentheses.

economically disadvantaged areas. To test this hypothesis, I introduce the interaction terms with public pawn loans and various variables representing poverty levels.

Table 6 presents the results of testing the heterogeneity associated with poverty levels. All estimating equations incorporate the same covariates as the baseline specification: all control variables, ward- and year-fixed effects, and ward-specific time trends.

Columns (1) and (5) include an interaction term for the loan amount from public pawnshops and a dummy variable denoting the poorest wards, namely Shimoya, Asakusa, Honjo, and Fukagawa. These wards are determined based on the income per capita in 1933 (Figure 2). In low-income areas, which likely experienced the poorest health, public pawn loans exhibit the most substantial relationship with health improvement. Columns (2) and (6) show the interaction terms for the amounts loaned by public pawnshops in years *t* and 1927. As public institutions are established in poorer areas, their loan amounts in 1927, the initial year of the study period, reflects the size of the population requiring financial assistance. I employ another proxy for the poverty level, the coverage of social workers, in columns (3) and (7). Although this variable has already been used as a control variable, its interaction with public pawn loans merits consideration, as it depends on the number of low-income people seeking help. Additionally, columns (4) and (8) incorporate an interaction term between public pawn loans and the ward-level poverty ratio. This measurement is the percentage of protection-requiring individuals, defined as those with income below the poverty line set by Tokyo, of the total population. As social surveys on protection-requiring individuals are not conducted annually, I use data from 1929, 1931, 1932, 1933, and 1935.[36]

The estimation results emphasize the homogeneity of the relationship between public pawn loans and favorable health conditions in economically disadvantaged and other wards. The estimated coefficients of public pawn loans remain negative and statistically significant in all equations, except for column (4), whereas the interaction terms between the loan amount and poverty level proxies are nonsignificant across all equations. These results reinforce my primary findings and suggest that the relationship between public pawn loans and health outcomes does not depend on the economic characteristics of wards, irrespective of the poverty level measurement used.

This finding is ambiguous because low-income households use public pawnshops. A possible explanation is that individuals who frequented public pawnshops have lower incomes if they live in affluent wards. Every ward, regardless of overall affluence or poverty level, likely has low-income residents who rely on pawnshops as financial resources. If ward-level infant mortality and fetal death rates largely depend on health conditions in the low-income class, public pawn loans assisting them can lead to health improvements, even in affluent wards.

[36] Data on protection-requiring individuals are obtained from surveys conducted by the Tokyo City Social Welfare Bureau (Tokyo City Office, 1930; 1932; 1933b; 1934b; Tokyo City Social Welfare Bureau, 1936).

Finally, I focus on the role of private pawnshops. The main findings highlight the differences in the contributions to health improvement between public and private pawnshops. Nonetheless, the intended use of loans from private institutions can vary, depending on the presence of their public counterparts. Thus, to explore the relationships between private pawn loans and health outcomes, I conduct further estimates based on a subsample of eight wards without public pawnshops, employing the same specifications as those in Table 2.

Table 7 displays the results, all of which are statistically nonsignificant, irrespective of the control variables. These results correspond with the primary results, supporting the marked distinction between the two types of pawnshops despite their small institutional differences. However, the estimated coefficients of private pawn loans shift in a negative direction for both infant and fetal death rates compared with those in Table 2. Hence, low-income households use private pawnshops for health-related purposes if there are no

Table 7: Relationships between Private Pawn Loans and Health Outcomes

| | Infant mortality rate | | | | Fetal death rate | | | |
|---|---|---|---|---|---|---|---|---|
| | (1) | (2) | (3) | (4) | (5) | (6) | (7) | (8) |
| Private loans | -0.506 | -0.495 | -0.551 | -0.537 | 0.003 | 0.043 | 0.019 | -0.010 |
| | (0.474) | (0.500) | (0.530) | (0.538) | (0.204) | (0.231) | (0.250) | (0.253) |
| Social worker | Yes | Yes | Yes | Yes | Yes | Yes | Yes | Yes |
| Taxpayer | No | Yes | Yes | Yes | No | Yes | Yes | Yes |
| Doctor | No | No | Yes | Yes | No | No | Yes | Yes |
| Water tap | No | No | No | Yes | No | No | No | Yes |
| Ward FE | Yes | Yes | Yes | Yes | Yes | Yes | Yes | Yes |
| Year FE | Yes | Yes | Yes | Yes | Yes | Yes | Yes | Yes |
| Time trend | Yes | Yes | Yes | Yes | Yes | Yes | Yes | Yes |
| Observations | 72 | 72 | 72 | 72 | 72 | 72 | 72 | 72 |

***, **, and * represent statistical significance at the 1%, 5%, and 10% levels, respectively.
Notes: These estimates use the data of wards without public pawnshops: Akasaka, Azabu, Hongo, Kanda, Koujimachi, Nihombashi, Ushigome, and Yotsuya. All regressions in columns (1)–(4) and (5)–(8) are weighted by the number of live births and that of births, respectively. The dependent variables and private pawn loans are logarithms. The loan amounts are the real values. Standard errors clustered at the ward level are in parentheses.

public pawnshops. This interpretation provides corroborative evidence that public pawnshops help economically disadvantaged people through lower interest loans.[37]

## 6. Discussion

This section provides approximate cost-effectiveness calculations for public pawnshops regarding mortality reduction to evaluate their role in public services. During the study period, the total cost of public pawnshops in Tokyo is 1,195,355 yen, of which approximately 35% is for facilities and 65% is for loans. Financial resources include earthquake relief funds, government subsidies, and borrowings from national and Tokyo prefectural governments (Tokyo Prefecture Department of Academic Affairs, 1935).[38] I do not consider annual costs such as employee salaries because these costs can be fully paid from annual income, including interest receipts.[39]

[37] If increasing throughout the period, wages of low-skilled workers might have decreased the mortality rates in poor wards and simultaneously correlated with the increase in public pawn loans, causing omitted variable bias. However, wages, including those of low-skilled workers such as day workers and dockworkers, remained stable from 1927 to 1934 (Figure A.4 in Appendices). I have further checked the robustness of the main estimation results in other various ways. See Appendix B for detail.

Setting 1927 as the reference year, I calculate the number of infants saved through public pawnshop loans between 1928 and 1935 using the following equation:

$$Saved\ infants = ID_{1927} \times \hat{\beta} \sum_{t=1928}^{1935} \left( \frac{Loan_t - Loan_{1927}}{Loan_{1927}} \right),$$

where $Loan_t$ denotes the total loan amount to public pawnshops in year *t*. The variable $ID_{1927}$ is the total number of infant deaths in wards with public pawnshops in 1927 (2,020), and $\hat{\beta}$ takes 0.022 according to the baseline result reported in column (4) of Table 2. The calculation results indicate that public pawnshops saved the lives of 1,169 infants during this period. Similarly, the number of avoided fetal deaths is 682.

---

[38] Table A.1 in Appendices displays financial resources and amounts in detail. The total cost includes expenditures for three pawnshops in the new city area.

[39] For example, in 1933, annual revenue is 1,505,647 yen, while annual expenditure is 1,295,719 yen. These are 69,991 yen and 69,374 yen, respectively, excluding loans, refunds, carryovers from the previous year, and bond redemptions (Tokyo City Social Welfare Bureau, 1935).

Based on the above cost and calculated number of saved infants, one additional infant can avoid death for every 1,023 yen (approximately $505 in 1930 dollars or $6,593 in 2010 dollars) spent on public pawnshops.[40] The cost of saving an infant is much less than that of the conditional cash transfer in the United States in the 1930s, $733,897 in 2019 dollars (Galofré 2020). Moreover, the infant saving cost of home nurse visits in other countries is higher, approximately $1,600 in the United States in 1927 and USD 10,314 (in 2010 dollars) in Denmark in 1941 (Moehling and Thomasson, 2014; Wüst, 2012). Thus, public pawnshops are a cost-effective social service for reducing infant mortality, even though health promotion is not their sole purpose.[41] This remarkably high efficiency is due to the consecutive interest receipts. The self-selection that only those who required money took out loans may have also kept the costs low.

Under the assumption that the daily wage was 2 yen, this study's estimates revealed that infants could earn the cost of saving their own lives through approximately 500 days of work in adulthood. Similarly, 1,753 yen spent on public pawnshops could prevent one fetus from dying, corresponding to a wage of roughly 900 days of work. These estimates emphasize high cost-effectiveness and imply that public pawnshops not only helped low-income people but also contributed to economic growth through mortality reduction.

[40] The conversion from the Japanese yen to the U.S. dollar is based on the exchange rate in 1930 (Bank of Japan 1986).

[41] Nevertheless, this calculation result was conservative because public pawnshops provided loans after 1937 without additional costs.

## 7. Conclusions

There is limited knowledge regarding the relationship between financial institutions for low-income individuals and the decline in mortality rates during the prewar period. However, the historical fact that pawnshops, particularly public ones, provided small loans at relatively low interest rates implies that these institutions improved the health of those facing financial challenges through easing liquidity constraints. Established with social welfare in mind, public pawnshops were broadly similar to their private counterparts but differed in interest rates and loan amounts. In Tokyo, public pawnshops charged less than half the interest rates of private pawnshops and their rates did not vary based on the loan amount, whereas private institutions charged higher interest rates for smaller loans. Consequently, public pawnshops benefitted low-income households, which frequently required smaller amounts of money. Furthermore, public pawnshops set upper limits on loan amounts to prevent middle-income individuals from availing loans, thereby focusing on offering loans to low-income individuals. These characteristics suggest that, compared to private pawnshops, public pawnshops supported access to credit among the poor more effectively and, as a result, promoted better health.

The empirical findings support this assertion. The estimation results indicated that public pawnshops contributed to reducing infant mortality and fetal death rates, whereas private pawn loans did not exhibit any relationship with health improvements. This disparity emanated from the income classes of pawnshop users. Low-income individuals, with poor health, primarily utilized public pawnshops. The loans received helped mitigate health-related risks by improving the nutritional status and hygiene. Consequently, loans from public pawnshops are significantly associated with health improvements. By contrast, middle-income individuals, who increasingly demand relatively large loans, frequent private pawnshops during that time (Social Welfare Bureau, 1926b; Shibuya et al., 1982). These individuals did not need to borrow small amounts of money for their living expenses; therefore, there was no clear association between the amount loaned by private pawnshops and health outcomes. The finding that only public pawn loans are related to health improvements implies that even small institutional changes designed to support economically disadvantaged individuals can substantially affect their well-being.

## References

Alsan, M., and Goldin, C. (2019). Watersheds in Child Mortality: The Role of Effective Water and Sewerage Infrastructure, 1880–1920. *The Journal of Political Economy,* 127(2), 586–638.

Anderson, D. M., Charles, K. K., and Rees, D. I. (2022). Reexamining the Contribution of Public Health Efforts to the Decline in Urban Mortality. *American Economic Journal. Applied Economics,* 14(2), 126–57.

Bank of Japan. (1913). *Shichiya ni kansuru chosa* [Survey on pawnbroking]. Tokyo.

Bank of Japan. (1986). *Nihon ginko hyakunenshi, shiryo hen* [The hundred-year history of the Bank of Japan, materials]. Tokyo.

Bauernschuster, S., Driva, A., and Hornung, E. (2020). Bismarck's Health Insurance and the Mortality Decline. *Journal of the European Economic Association,* 18(5), 2561–607.

Bhalotra, S., Karlsson, M., and Nilsson, T. (2017). Infant Health and Longevity: Evidence from A Historical Intervention in Sweden. *Journal of the European Economic Association,* 15(5), 1101–57.

Bhutta, N. (2014). Payday Loans and Consumer Financial Health. *Journal of Banking & Finance,* 47, 230–42.

Bowblis, J. R. (2010). The decline in infant death rates, 1878–1913: the role of early sickness insurance programs. *Journal of Economic History,* 70(1), 221–232.

Bureau of Social Affairs. (1922). *Saiminchosa tokeihyo* [Statistical tables of the conditions of the low-income households]. Tokyo.

Cutler, D., and Miller, G. (2005). The role of public health improvements in health advances: the twentieth-century United States. *Demography,* 42(1), 1–22.

Ferrie, J., and Troesken, W. (2008). Water and Chicago's Mortality Transition, 1850–1925. *Explorations in Economic History,* 45(1), 1–16.

Galofré Vilà, G. (2020). Quantifying the Impact of Aid to Dependent Children: An Epidemiological Framework. *Explorations in Economic History,* 77, 101332.

Goldenberg, R. L., Elizabeth, M. M., Sarah, S., and Uma M. R. (2010). Infection-Related Stillbirths. *The Lancet,* 375(9724), 1482–90.

Goldin, C. (2024). Human Capital. In: Diebolt, C., and Haupert, M. (Eds.) *Handbook of Cliometrics* (3rd ed.). Switzerland: Springer Nature.

Horrell, S., and Deborah, O. (2000). Work and Prudence: Household Responses to Income Variation in Nineteenth-Century Britain. *European Review of Economic History,* 4(1), 27–57.

Inoue, T. (2021). The Role of Pawnshops in Risk Coping in Early Twentieth-Century Japan. *Financial History Review,* 28(3), 319–43.

Inoue, T., and Ogasawara, K. (2020). Chain Effects of Clean Water: The Mills-Reincke Phenomenon in Early 20th-Century Japan. *Economics and Human Biology,* 36, 100822.

Ioku, S., and Suizume, M. (2014). Kindai nihon no shomin kinyu: Tokyoshi Shibaku T shichiten no kenkyu [Finance for common people in modern Japan: the case of "T" pawnshop in Shiba ward in Tokyo]. *Shakai Keizaishi Gaku,* 80(3), pp. 3–8.

James, J. A., and Isao, S. (2011). Early Twentieth-Century Japanese Worker Saving: Precautionary Behaviour before a Social Safety Net. *Cliometrica,* 5(1), 1–25.

Kagawa, T. (1915). *Himmin shinri no kenkyu* [A Study of the Psychology of the Poor]. Tokyo: Keiseisha.

Kao Museum and Material Room (eds). (2012). *Kao 120 nen, 1890–2010 nen* [Kao 120 years, 1890–2010]. Tokyo.

Kawabata, M. (2016). *Kindai nihon no koshuyokujo undo.* [Public Bath Movement in Modern Japan]. Tokyo: Hosei Daigaku Shuppamkyoku.

Kiesling, L. L. (1996). Institutional Choice Matters: The Poor Law and Implicit Labor Contracts in Victorian Lancashire. *Explorations in Economic History,* 33(1), 65–85.

Kojima, Y. (2021). *Sarakin no rekishi: Shohisha kinyu to nihon shakai*. [History of Salaryman loan: Consumer loan and Japanese society]. Tokyo: Chukoron-shinsha.

McLaughlin, E., and Rowena, P. (2022). Fringe Banking and Financialization: Pawnbroking in Pre-famine and Famine Ireland. *The Economic History Review,* 75(3), 903–31.

Miyamoto, M. (2008). Kindai keizai seicho [Modern economic growth]. In: M. Miyamoto (ed), *Nihon keizaishi* [Japanese Economic History] (pp. 53–111). Tokyo: Hosodaigaku kyoiku shinkokai.

Moehling, C. M., and Thomasson, M. A. (2014). Saving babies: the impact of public education programs on infant mortality. *Demography*, 51(2), 367–86.

Murhem, S. (2016). Credit for the Poor: The Decline of Pawnbroking 1880–1930. *European Review of Economic History,* 20(2), 198–214.

Murhem, S., and Ulväng, G. (2023). Pawning and pawners in the industrial era: evidence from Sweden, 1870 to 1950. *History of Retailing and Consumption*, 9(1), 38–56.

Ogasawara, K. (2022). Persistence of Natural Disasters on Children's Health: Evidence from the Great Kantō Earthquake of 1923. *The Economic History Review,* 75(4), 1054–82.

Ogasawara, K. (2024a). Cliometrics of Child Health. In: Diebolt, C., and Haupert, M. (Eds.) *Handbook of Cliometrics* (3rd ed.). Switzerland: Springer Nature.

Ogasawara, K. (2024b). Consumption Smoothing in the Working-Class Households of Interwar Japan. *The Journal of Economic History* 84 (1), 111–48.

Ogasawara, K., and Kobayashi, G. (2015). The Impact of Social Workers on Infant Mortality in Inter-War Tokyo: Bayesian Dynamic Panel Quantile Regression with Endogenous Variables. *Cliometrica,* 9(1), 97–130.

Ogasawara, K., and Matsushita, Y. (2018). Public Health and Multiple-Phase Mortality Decline: Evidence from Industrializing Japan. *Economics and Human Biology,* 29, 198–210.

Ogasawara, K., and Matsushita, Y. (2019). Heterogeneous Treatment Effects of Safe Water on Infectious Disease: Do Meteorological Factors Matter? *Cliometrica,* 13(1), 55–82.

Ogasawara, K., Shirota, S., and Kobayashi, G. (2018). Public Health Improvements and Mortality in Interwar Tokyo: A Bayesian Disease Mapping Approach. *Cliometrica,* 12(1), 1–31.

Ogasawara, S. (1913). *Shichiya no kenkyu* [A Study on Pawnshops]. Tokyo: Ryomeidoshoten.

Okazaki, T., Toshihiro, O., and Eric, S. (2019). Creative Destruction of Industries: Yokohama City in the Great Kanto Earthquake, 1923. *The Journal of Economic History,* 79(1), 1–31.

Osaka City Social Welfare Bureau. (1942). *Shomin kinyu jijo chosa* [Survey on the monetary situation of common people]. Osaka.

Saaritsa, S. (2008). Informal Transfers, Men, Women and Children: Family Economy and Informal Social Security in Early 20th Century Finnish Households. *The History of the Family,* 13(3), 315–31.

Saaritsa, S. (2011). The Poverty of Solidarity: The Size and Structure of Informal Income Smoothing among Worker Households in Helsinki, 1928. *The Scandinavian Economic History Review,* 59(2), 102–27.

Santaya, H. (1941). *Nyuyoji hogo* [Saving babies and infants]. Tokyo: Dobunkan.

Scott, P. M., and James, W. (2012). Working-Class Household Consumption Smoothing in Interwar Britain. *The Journal of Economic History,* 72(3), 797–825.

Shaffer, R. (2013). 'A Missionary from the East to Western Pagans': Kagawa Toyohiko's 1936 U.S. Tour. *Journal of World History,* 24(3), 577–621.

Shibuya, R, Suzuki, K, Ishiyama, S. (1982). *Nihon no shichiya* [Pawnshop in Japan]. Tokyo: Waseda Daigaku Shuppambu.

Schneider, Eric B. (2025). "The Determinants of Child Stunting and Shifts in the Growth Pattern of Children: A Long-run, Global Review." *Journal of Economic Surveys* 39 (2), 405–52.

Social Welfare Bureau. (1926a). *Koeki shichiko no genzei* [Current situation of public pawnshops]. Tokyo.

Social Welfare Bureau. (1926b). *Tokyoshinai oyobi gumbu ni okeru shichiya ni kansuru chosa* [Survey on pawnshops in both city and county in Tokyo]. Tokyo.

Social Welfare Bureau. (1926c). *Tokyoshi koeki shichiko ni kansuru chosa* [Survey on public pawnshops in Tokyo City]. Tokyo.

Statistical Division of the Commerce and Industry Minister. (various years). *Oroshiuri bukka tokeihyo* [Statistics of wholesale price index]. Tokyo.

Tokyo Chamber of Commerce and Industry. (1935). *Tokyo shokokaigisho tokei nempo, showa 10nen* [Annual statistical report of the Tokyo Chamber of Commerce and Industry, 1935]. Tokyo.

Tokyo City. (1929–1937). *Tokyoshi tokei nempyo* [Statistical Yearbooks of Tokyo City, volumes 25–33]. Tokyo.

Tokyo City Office. (1926). *Shichiyagyo no tokei chosa* [Statistical survey on pawnbroking]. Tokyo.

Tokyo City Office. (1930). *Tokyoshinai yohogosha ni kansuru chosa* [Survey of protection-required people in Tokyo City]. Tokyo.

Tokyo City Office. (1932). *Tokyoshi yohogosetai seikei chosa* [Living survey of protection-required households in Tokyo City]. Tokyo.

Tokyo City Office. (1933a). *Tokyoshimin no shotoku chosa* [Income survey on citizen of Tokyo]. Tokyo.

Tokyo City Office. (1933b). *Tokyoshi yohogosha chosa* [Survey of protection-required people in Tokyo City]. Tokyo.

Tokyo City Office. (1934). *Tokyoshi yohogosetai chosa* [Survey of protection-required households in Tokyo City]. Tokyo.

Tokyo City Office. (1937). *Yohogosetai ni okeru shussho shizan narabini nyuyoji shibo jijo ni kansuru chosa* [Survey on the live births, deaths births, and infant mortality in the low-income households]. Tokyo.

Tokyo City Social Welfare Bureau. (1926, 1928–1936). *Tokyoshi shakaikyoku nempo* [Annual Reports of the Tokyo City Social Welfare Bureau, volumes 6, 8–16]. Tokyo.

Tokyo City Social Welfare Bureau. (1926b). *Nyuji shiboritsu ni kansuru chosa* [Survey on Infant Mortality Rate]. Tokyo.

Tokyo City Social Welfare Bureau. (1930b–1932b). *Tokyo shinai shibou narabi shizan ni kansuru chosa* [Survey on mortality and fetal deaths in Tokyo City, 1930–1932 editions]. Tokyo.

Tokyo City Social Welfare Bureau. (1932c). Current situation of poorly fed children in city [Shinai kesshoku jido no genjo]. In: Printing Bureau, Ministry of Finance (ed), *Official Gazette*, vol. 1543, Appendix [Kampo 1543go furoku zappo] (p. 4). Tokyo.

Tokyo City Social Welfare Bureau. (1934b). *Tokyoshi yohogosetai gaiyochosa* [Survey of protection-required households]. Tokyo.

Tokyo City Social Welfare Bureau. (1935a). *Hikyugosha ni kansuru chosa* [Survey of aided people]. Tokyo.

Tokyo City Social Welfare Bureau. (1935b). *Yohogosetai ni okeru nyuyoji no seikatsu jotai* [Living conditions of infants and children in protection-requiring households]. Tokyo.

Tokyo Institute for Municipal Research. (1926). *Kosetsu shichiho* [Public pawnshops]. Tokyo.

Tokyo Institute for Municipal Research. (1928). *Toshi ni okeru ninsampu hogo jigyo* [Maternal aid programs in cities]. Tokyo.

Tokyo Pawnshop Association. (1934). *Shichiya rishi no kenkyu* [Study of interest rate of pawnshops]. Tokyo: Tokyo Shichiya Kumiai.

Tokyo Prefecture. (1926–1937). *Tokyofu tokeisho* [Statistical Yearbooks of Tokyo Prefecture, 1924–1935 editions]. Tokyo.

Tokyo Prefecture Department of Academic Affairs. (1935). *Saimin kinyu ni kansuru chosa* [Survey on financial climate of the poor]. Tokyo.

Winegarden, C. R., and Murray, J. E. (1998). The Contributions of Early Health-Insurance Programs to Mortality Declines in Pre-World War I Europe: Evidence from Fixed-Effects Models. *Explorations in Economic History,* 35(4), 431–46.

Winegarden, C. R., and Murray, J. E. (2004). Effects of Early Health-Insurance Programs on European Mortality and Fertility Trends. *Social Science & Medicine,* 58(10), 1825–36.

Wüst, M. (2012). Early Interventions and Infant Health: Evidence from the Danish Home Visiting Program. *Labour Economics,* 19(4), 484–95.

# Appendices

Appendix A: Figures and Tables

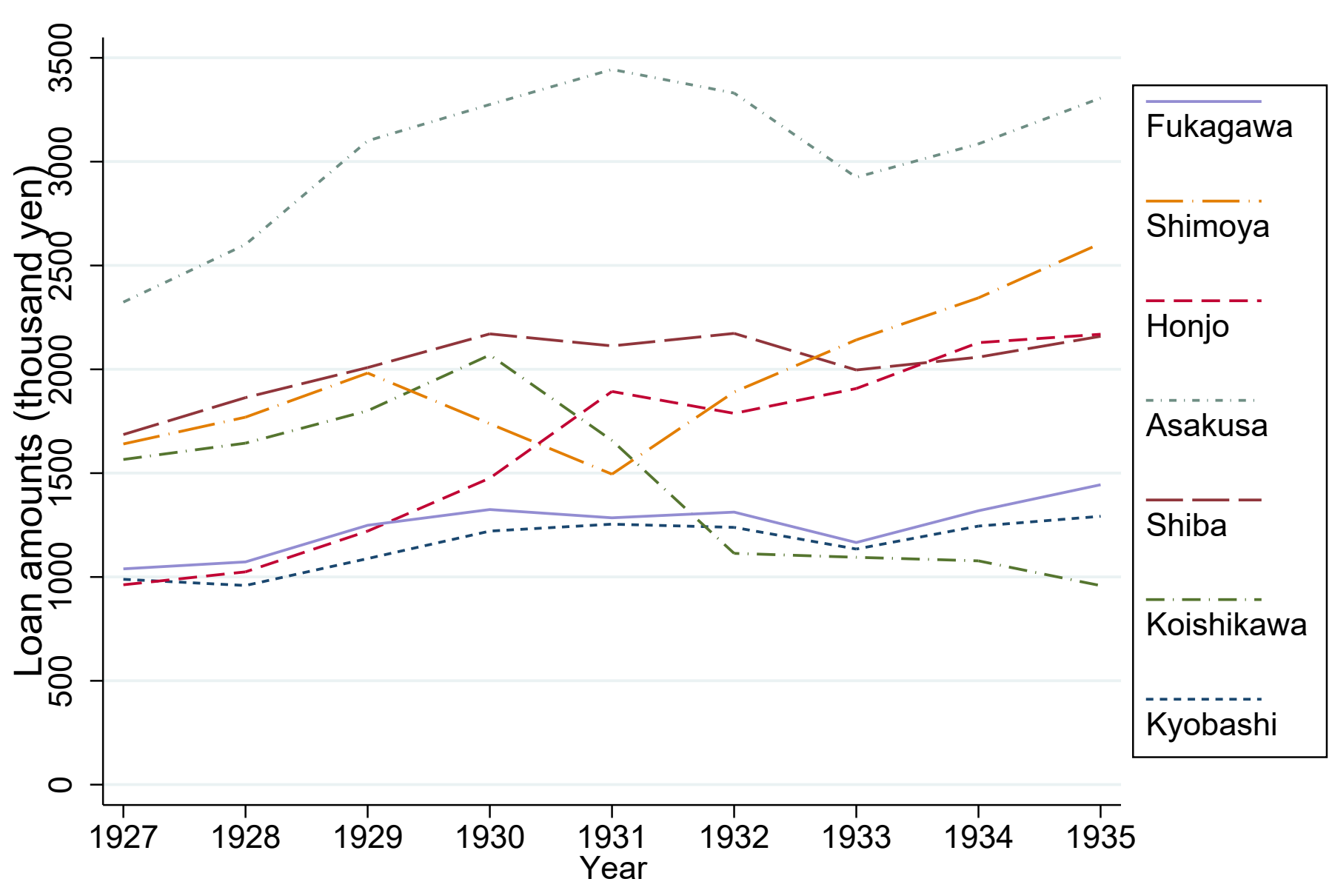

(a) Wards with Public Pawnshops

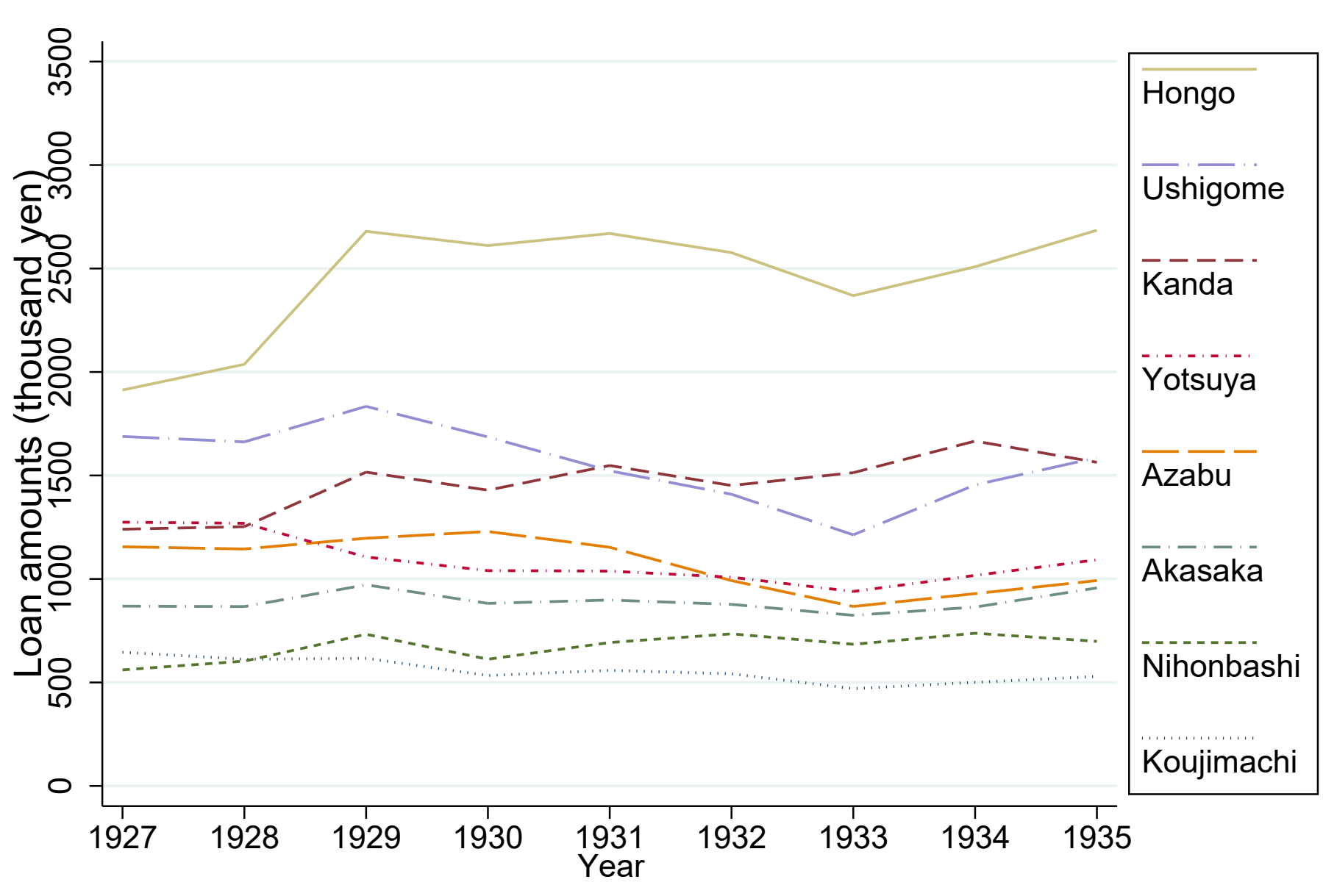

(b) Wards without Public Pawnshops

Figure A.1: Loan Amounts of Private Pawnshops

Note: Loan amounts are the real values adjusted by the wholesale price index based on December 1929. The order of the legend for wards with public pawnshops is the same as in Figure 3.

Sources: Tokyo City (1929–1937).

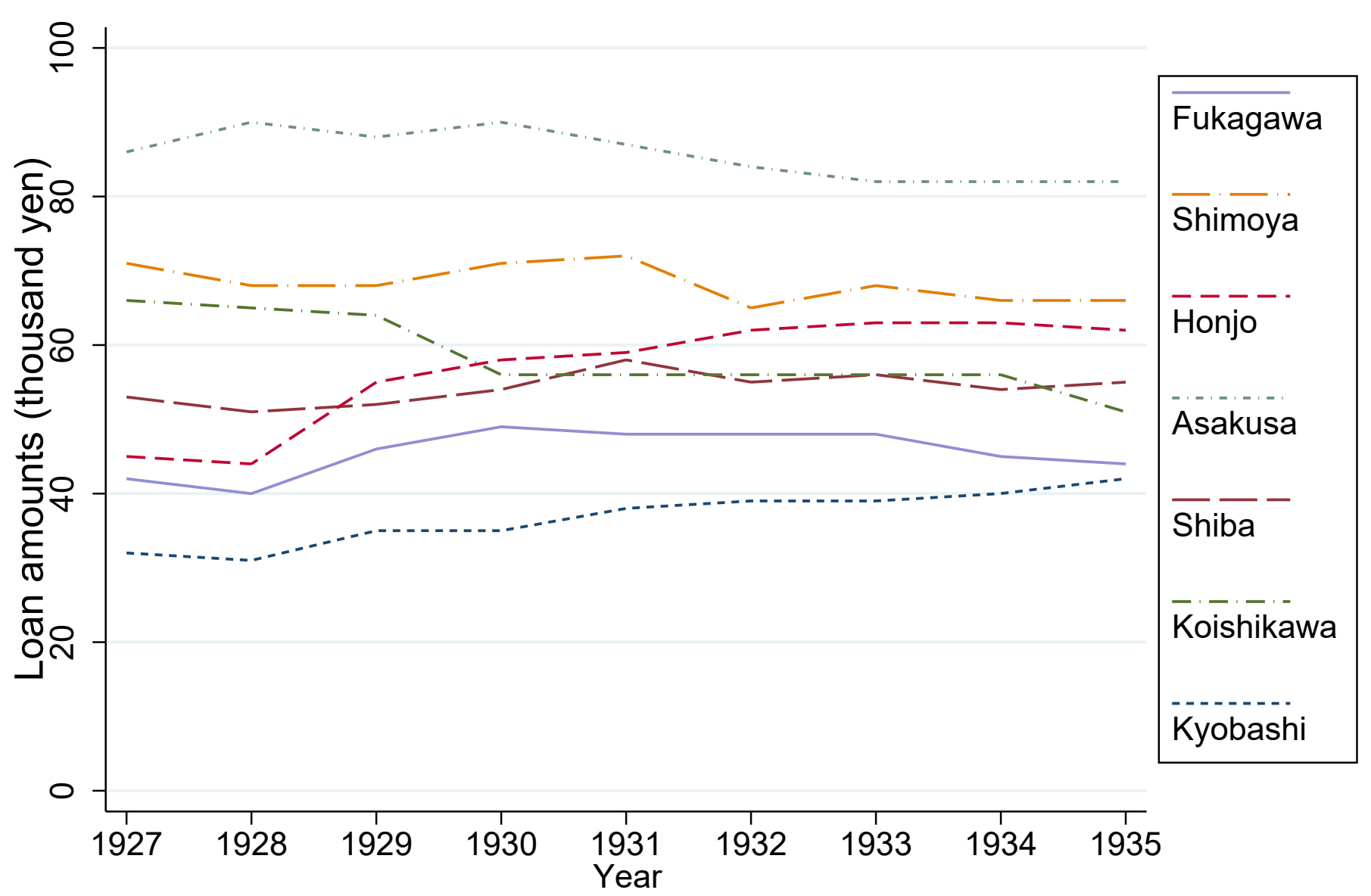

(a) Wards with Public Pawnshops

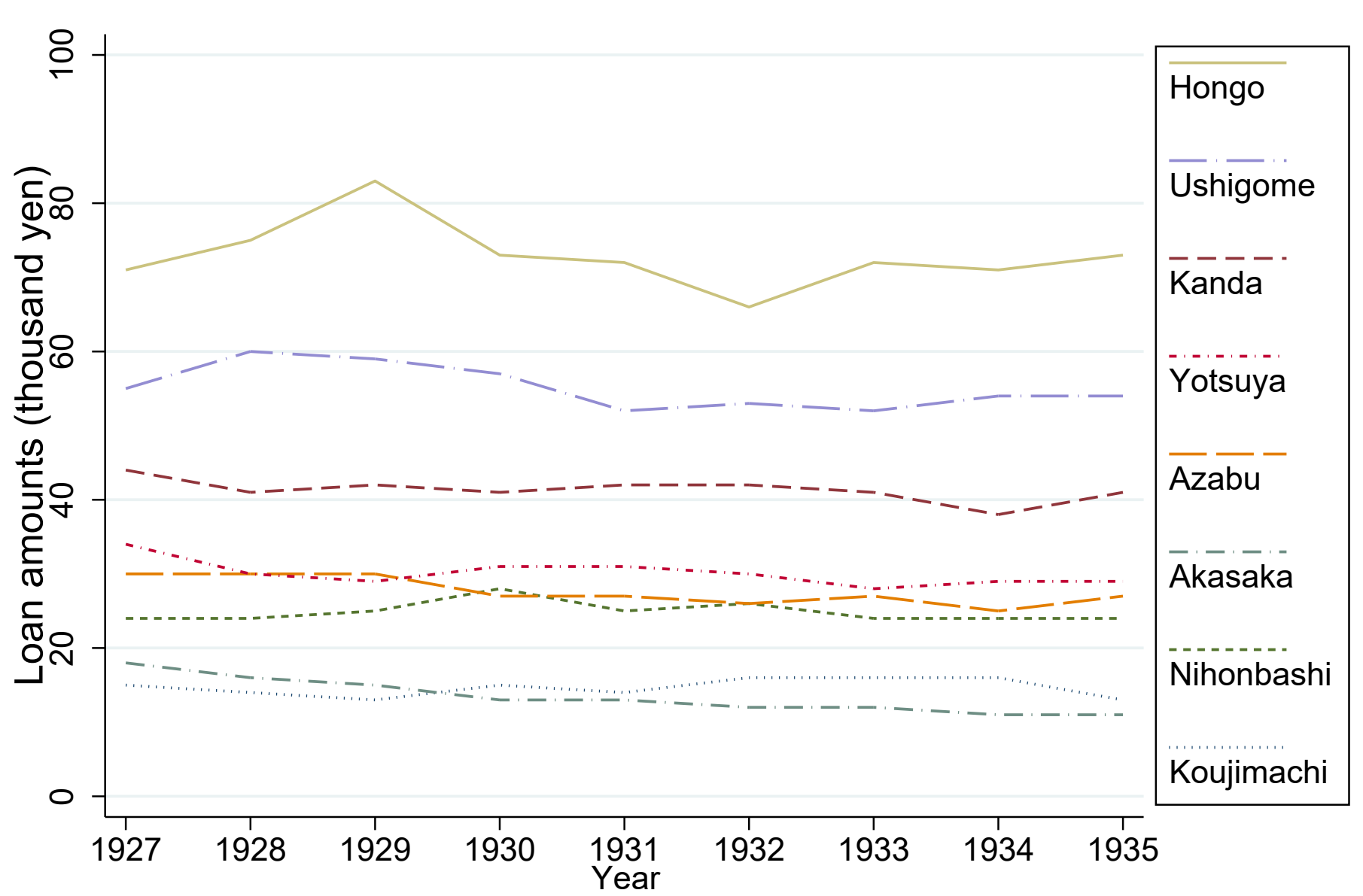

(b) Wards without Public Pawnshops

Figure A.2: Number of Private Pawnshops

Note: The order of the legend for wards with and without public pawnshops is the same as in Figures 3 and A.1, respectively.

Sources: Tokyo City (1929–1937).

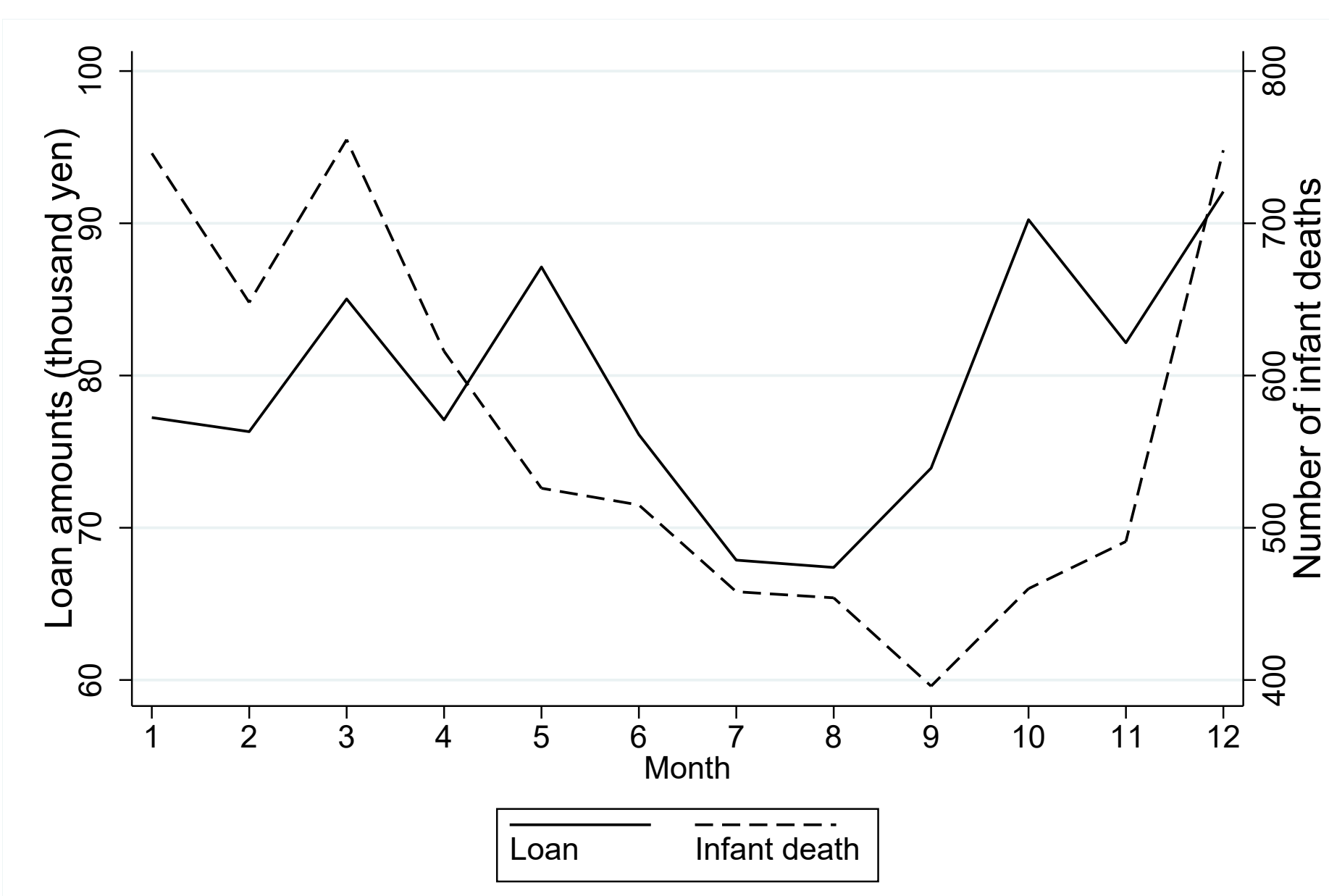

Figure A.3: Seasonality of Loan Amounts of Public Pawnshops and Infant Deaths

Note: The loan amounts are the real values adjusted by the wholesale price index based on December 1929.

Sources: Health Bureau of the Metropolitan Police Department (1928); Tokyo City Social Welfare Bureau (1928–1935).

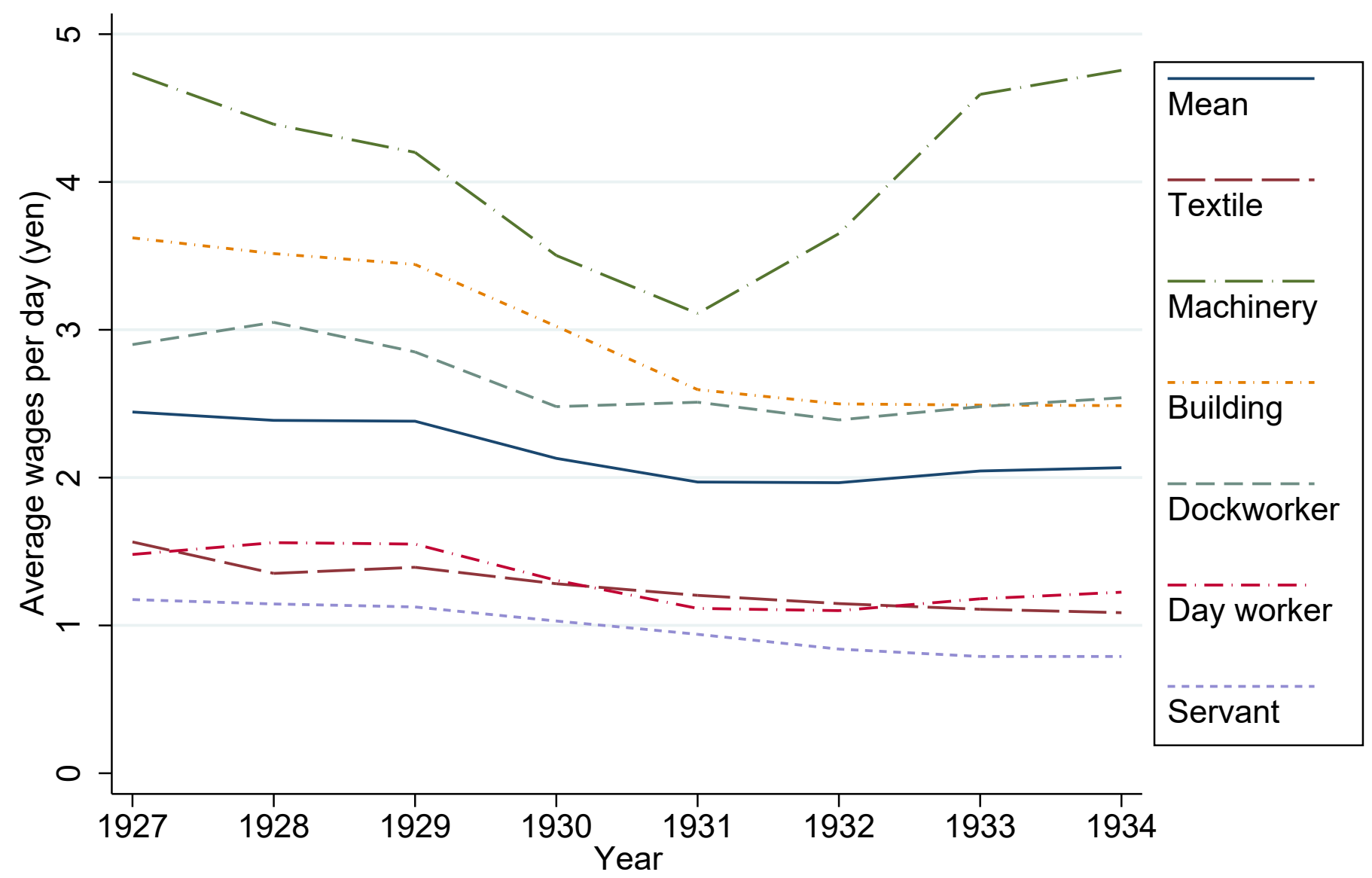

(a) Nominal Wages

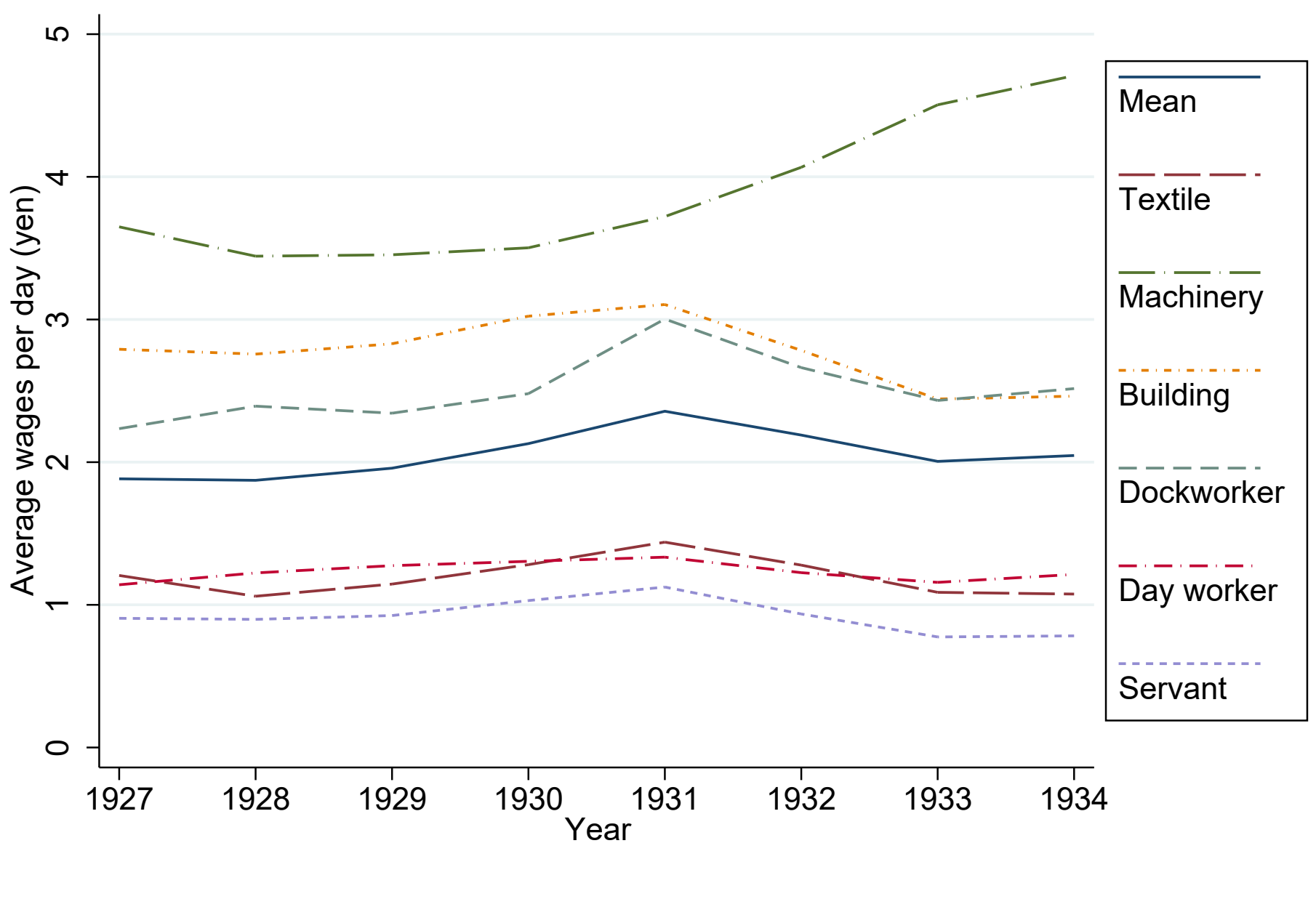

(b) Real Wages

Figure A.4: Average Wages per Day by Industries and Occupations in Tokyo

Note: The year 1935 is missing due to unavailability of consistent data. The average wages are based on wages in the six industries and occupations displayed and all other labor wages surveyed. The real wages are adjusted by the wholesale price index based on 1930.

Sources: Tokyo Chamber of Commerce and Industry (1935, pp.17, 27–31).

Table A.1: Financial Resources for Public Pawnshops in Tokyo City

| Year | Resource | Amount (yen) | | | Type |
|---|---|---|---|---|---|
| | | Total | Facility | Loan | |
| 1923 | Earthquake relief funds | 150,000 | 19,751 | 130,249 | Donation |
| 1925 | Home Ministry | 200,000 | 39,804 | 160,196 | Subsidy |
| 1926 | Home Ministry | 420,000 | 200,000 | 220,000 | Subsidy |
| 1930 | Tokyo Communication Department | 37,000 | 5,702 | 31,298 | Borrowing |
| 1931 | Tokyo Communication Department | 34,400 | 5,400 | 29,000 | Borrowing |
| 1931 | Finance Ministry | 210,000 | 0 | 210,000 | Borrowing |
| 1925–1930 | Earthquake reconstruction funds | 143,955 | 143,955 | 0 | Borrowing |

Sources: Tokyo Prefecture Department of Academic Affairs (1935).

Appendix B: Additional robustness checks

Appendix B.1: Estimates Based on Subsample Excluding Initial and Final Years

Table B.1: Relationships between Pawn Loans and Health Outcomes in 1928–1934

| | Infant mortality rate | | | | Fetal death rate | | | |
|---|---|---|---|---|---|---|---|---|
| | (1) | (2) | (3) | (4) | (5) | (6) | (7) | (8) |
| Public loans | -0.029*** | -0.026*** | -0.025*** | -0.021** | -0.033*** | -0.032*** | -0.031*** | -0.023*** |
| | (0.009) | (0.008) | (0.008) | (0.009) | (0.006) | (0.006) | (0.005) | (0.005) |
| Private loans | 0.043 | -0.020 | -0.016 | -0.032 | 0.251 | 0.232 | 0.247 | 0.216 |
| | (0.180) | (0.144) | (0.145) | (0.146) | (0.210) | (0.201) | (0.202) | (0.216) |
| Social worker | Yes | Yes | Yes | Yes | Yes | Yes | Yes | Yes |
| Taxpayer | No | Yes | Yes | Yes | No | Yes | Yes | Yes |
| Doctor | No | No | Yes | Yes | No | No | Yes | Yes |
| Water tap | No | No | No | Yes | No | No | No | Yes |
| Ward FE | Yes | Yes | Yes | Yes | Yes | Yes | Yes | Yes |
| Year FE | Yes | Yes | Yes | Yes | Yes | Yes | Yes | Yes |
| Time trend | Yes | Yes | Yes | Yes | Yes | Yes | Yes | Yes |
| Observations | 105 | 105 | 105 | 105 | 105 | 105 | 105 | 105 |

***, **, and * represent statistical significance at the 1%, 5%, and 10% levels, respectively.
Notes: All regressions in columns (1)–(4) and (5)–(8) are weighted by the number of live births and that of births, respectively. The dependent variables and private pawn loans are logarithms, while the inverse hyperbolic sine transformation is applied to public pawn loans. The loan amounts are the real values. Standard errors clustered at the ward level are in parentheses.

As the loan amounts in 1927 and 1935 are estimated based on those in some months of the years, this study constructs a subsample excluding these years to check their influences on the estimates. The results estimated using the data in 1928–1934 are reported in Table B.1. The specifications are the same as Table 2 (main text) except for the observations. Regardless of outcomes, the estimated coefficients of public pawn loan amounts are negative and statistically significant, while the effects of private pawn loans are insignificant. These results prove that the estimated loan amounts for 1927 and 1935 have no substantial influence on our regression analyses.

Appendix B.2: Tests Varying the Units of Inverse Hyperbolic Sine Function

Table B.2: Estimates Using Pawn Loans Measured by Different Units

| | Infant mortality rate | | | Fetal death rate | | |
|---|---|---|---|---|---|---|
| | (1) | (2) | (3) | (4) | (5) | (6) |
| Public loans | -0.027** | -0.034** | -0.045* | -0.036*** | -0.047*** | -0.066*** |
| | (0.009) | (0.013) | (0.022) | (0.008) | (0.012) | (0.020) |
| Private loans | -0.042 | -0.040 | -0.040 | 0.213 | 0.216 | 0.219 |
| | (0.164) | (0.164) | (0.163) | (0.183) | (0.180) | (0.175) |
| All control variables | Yes | Yes | Yes | Yes | Yes | Yes |
| Ward FE | Yes | Yes | Yes | Yes | Yes | Yes |
| Year FE | Yes | Yes | Yes | Yes | Yes | Yes |
| Time trend | Yes | Yes | Yes | Yes | Yes | Yes |
| Loan unit | 10 yen | 100 yen | 1,000 yen | 10 yen | 100 yen | 1,000 yen |
| Observations | 135 | 135 | 135 | 135 | 135 | 135 |

***, **, and * represent statistical significance at the 1%, 5%, and 10% levels, respectively.
Notes: All regressions in columns (1)–(4) and (5)–(8) are weighted by the number of live births and that of births, respectively. The dependent variables and private pawn loans are logarithms, while the inverse hyperbolic sine transformation is applied to public pawn loans. The loan amounts are the real values. All control variables are social worker, taxpayer, doctor, and water tap rates. Standard errors clustered at the ward level are in parentheses.

The estimated coefficients of public pawn loans can vary according to the measurement unit of loan amount because the variable is an inverse hyperbolic sine function. Thus, this study confirms that the main findings do not depend on any particular unit by changing the measurement unit of loan amounts in the baseline estimating equations. Table B.2 shows the results with different units, which are 10, 100, and 1,000 yen in columns (1) and (4), (2) and (5), and (3) and (6), respectively. All results are significantly negative, indicating that the study objective has been achieved. The magnitudes of the estimated coefficients are larger than those of the main results, suggesting that the primary analyses provide conservative estimates.

Appendix B.3: Estimates with Different Measures

Table B.3: Estimates Using Loan Amounts per Households or Population

| | Infant mortality rate | | Fetal death rate | |
|---|---|---|---|---|
| | (1) | (2) | (3) | (4) |
| Panel A: per 1,000 households | | | | |
| Public loans | -0.030** | | -0.041*** | |
| | (0.011) | | (0.010) | |
| Private loans | | -0.042 | | 0.161 |
| | | (0.151) | | (0.161) |
| Panel B: per 1,000 people | | | | |
| Public loans | -0.035** | | -0.051*** | |
| | (0.015) | | (0.013) | |
| Private loans | | -0.064 | | 0.204 |
| | | (0.170) | | (0.187) |
| All control variables | Yes | Yes | Yes | Yes |
| Ward FE | Yes | Yes | Yes | Yes |
| Year FE | Yes | Yes | Yes | Yes |
| Time trend | Yes | Yes | Yes | Yes |
| Observations | 135 | 135 | 135 | 135 |

***, **, and * represent statistical significance at the 1%, 5%, and 10% levels, respectively.
Notes: All regressions in columns (1)–(4) and (5)–(8) are weighted by the number of live births and that of births, respectively. The dependent variables and private pawn loans are logarithms, while the inverse hyperbolic sine transformation is applied to public pawn loans. The loan amounts are the real values. All control variables are social worker, taxpayer, doctor, and water tap rates. Standard errors clustered at the ward level are in parentheses.

To verify that the results are robust regardless of the measures of the loans, this study conducts sensitivity tests using public and private pawn loans per 1,000 households or 1,000 people, the number of pawnshop users did not necessarily increase with population growth. The results using loan amounts per 1,000 households instead of the absolute amounts are presented in Panel A of Table B.3, while Panel B uses loans per 1,000 people. The estimated coefficients align with the main results and exhibit stronger relationships to health outcomes despite the same specifications. Thus, these estimates support that the primary findings are not due to arbitrary measures of the key variable.

Appendix B.4: Unweighted Regression Analyses

Table B.4: Estimation Results of Unweighted Regressions

| | Infant mortality rate | | | | Fetal death rate | | | |
|---|---|---|---|---|---|---|---|---|
| | (1) | (2) | (3) | (4) | (5) | (6) | (7) | (8) |
| Public loans | -0.021** | -0.022** | -0.022** | -0.021** | -0.032*** | -0.032*** | -0.031*** | -0.028*** |
| | (0.009) | (0.008) | (0.008) | (0.008) | (0.006) | (0.006) | (0.006) | (0.007) |
| Private loans | -0.085 | -0.113 | -0.115 | -0.117 | 0.116 | 0.102 | 0.091 | 0.079 |
| | (0.192) | (0.196) | (0.197) | (0.200) | (0.181) | (0.177) | (0.179) | (0.188) |
| Social worker | Yes | Yes | Yes | Yes | Yes | Yes | Yes | Yes |
| Taxpayer | No | Yes | Yes | Yes | No | Yes | Yes | Yes |
| Doctor | No | No | Yes | Yes | No | No | Yes | Yes |
| Water tap | No | No | No | Yes | No | No | No | Yes |
| Ward FE | Yes | Yes | Yes | Yes | Yes | Yes | Yes | Yes |
| Year FE | Yes | Yes | Yes | Yes | Yes | Yes | Yes | Yes |
| Time trend | Yes | Yes | Yes | Yes | Yes | Yes | Yes | Yes |
| Observations | 135 | 135 | 135 | 135 | 135 | 135 | 135 | 135 |

***, **, and * represent statistical significance at the 1%, 5%, and 10% levels, respectively.
Notes: The dependent variables and private pawn loans are logarithms, while the inverse hyperbolic sine transformation is applied to public pawn loans. The loan amounts are the real values. Standard errors clustered at the ward level are in parentheses.

Despite employing the regression analyses weighted by the mean values of the denominators of the dependent variable as the baseline identification strategy, this study checks the robustness of the results by running unweighted regressions. The estimated coefficients reported in Table B.4 are very similar to the baseline results, thus indicating that the estimation results remain stable irrespective of weighting.

Appendix B.5: Tests Using Nominal Loan Amounts

Table B.5: Estimates Using Nominal Values of Pawnshop Loan Amounts

| | Infant mortality rate | | | | Fetal death rate | | | |
|---|---|---|---|---|---|---|---|---|
| | (1) | (2) | (3) | (4) | (5) | (6) | (7) | (8) |
| Public loans | -0.025*** | -0.024*** | -0.024*** | -0.023*** | -0.035*** | -0.035*** | -0.034*** | -0.032*** |
| | (0.008) | (0.007) | (0.007) | (0.007) | (0.006) | (0.005) | (0.005) | (0.007) |
| Private loans | -0.016 | -0.037 | -0.037 | -0.043 | 0.238 | 0.228 | 0.225 | 0.211 |
| | (0.155) | (0.159) | (0.159) | (0.165) | (0.186) | (0.179) | (0.177) | (0.185) |
| Social worke | Yes | Yes | Yes | Yes | Yes | Yes | Yes | Yes |
| Taxpayer | No | Yes | Yes | Yes | No | Yes | Yes | Yes |
| Doctor | No | No | Yes | Yes | No | No | Yes | Yes |
| Water tap | No | No | No | Yes | No | No | No | Yes |
| Ward FE | Yes | Yes | Yes | Yes | Yes | Yes | Yes | Yes |
| Year FE | Yes | Yes | Yes | Yes | Yes | Yes | Yes | Yes |
| Time trend | Yes | Yes | Yes | Yes | Yes | Yes | Yes | Yes |
| Observations | 135 | 135 | 135 | 135 | 135 | 135 | 135 | 135 |

***, **, and * represent statistical significance at the 1%, 5%, and 10% levels, respectively.
Notes: All regressions in columns (1)–(4) and (5)–(8) are weighted by the number of live births and that of births, respectively. The dependent variables and private pawn loans are logarithms, while the inverse hyperbolic sine transformation is applied to public pawn loans. The loan amounts are the nominal values. Standard errors clustered at the ward level are in parentheses.

In the regression analyses, the loan amounts of pawnshops are the real values adjusted by the wholesale price index based on December 1929. This seems reasonable; nonetheless, the validity of the primary findings is corroborated by using nominal values of loan amounts instead. The estimation results using nominal loan amounts are presented in Table B.5. Obviously, the estimated coefficients of public and private pawnshop loans are largely unchanged. This indicates that the conversion from nominal to real values have no influence on the estimated relationships between pawn loans and health outcomes.

References

Health Bureau of the Metropolitan Police Department. 1928. *Nyuyoji shibo chosa hokoku, taisho 15 nen showa gan nen* [Report of survey on infant and child deaths, 1925–6], Tokyo.

Tokyo Chamber of Commerce and Industry. 1935. *Tokyo shokokaigisho tokei nempo, showa 10nen* [Annual statistical report of the Tokyo Chamber of Commerce and Industry, 1935], Tokyo.

Tokyo Prefecture Department of Academic Affairs. 1935. *Saimin kinyu ni kansuru chosa* [Survey on financial climate of the poor], Tokyo.